\documentclass[twocolumn,  tighten, twocolappendix]{aastex63}

\shorttitle{Scatter in the Fundamental Plane of Early-type Galaxies}
\shortauthors{Yoon \& Park}

\begin{document}

\title{Dependence of the Fundamental Plane of Early-type Galaxies on Age and Internal Structure}

\email{yyoon@kias.re.kr}

\author[0000-0003-0134-8968]{Yongmin Yoon}
\affiliation{School of Physics, Korea Institute for Advanced Study (KIAS), 85 Hoegiro, Dongdaemun-gu, Seoul, 02455, Republic of Korea}

\author{Changbom Park}
\affiliation{School of Physics, Korea Institute for Advanced Study (KIAS), 85 Hoegiro, Dongdaemun-gu, Seoul, 02455, Republic of Korea}

\begin{abstract}
We investigate the scatter in the fundamental plane (FP) of early-type galaxies (ETGs) and its dependence on age and internal structure of ETGs, using $16,283$ ETGs with $M_r\le-19.5$ and $0.025\le z<0.055$ in Sloan Digital Sky Survey data. We use the relation between the age of ETGs and photometric parameters such as color, absolute magnitude, and central velocity dispersion of ETGs and find that the scatter in the FP depends on age. The FP of old ETGs with age $\gtrsim9$ Gyrs has a smaller scatter of $\sim0.06$ dex ($\sim14\%$) while that of young ETGs with age $\lesssim6$ Gyrs has a larger scatter of  $\sim0.075$ dex ($\sim17\%$). In the case of young ETGs, less compact ETGs have a smaller scatter in the FP ($\sim0.065$ dex; $\sim15\%$) than more compact ones ($\sim0.10$ dex; $\sim23\%$). On the other hand, the scatter in the FP of old ETGs does not depend on the compactness of galaxy structure. Thus, among the subpopulations of ETGs, compact young ETGs have the largest scatter in the FP. This large scatter in compact young ETGs is caused by ETGs that have low dynamical mass-to-light ratio ($M_\mathrm{dyn}/L$) and blue color in the central regions. By comparing with a simple model of the galaxy that has experienced a gas-rich major merger, we find that the scenario of recent gas-rich major merger can reasonably explain the properties of the compact young ETGs with excessive light for a given mass (low $M_\mathrm{dyn}/L$) and blue central color.
\end{abstract}

\keywords{galaxies: elliptical and lenticular, cD  --- galaxies: fundamental parameters --- galaxies: structure}

\section{Introduction}\label{sec:intro}

Early-type galaxies (ETGs) are virialized systems, so that they are expected to satisfy the balance between potential and kinetic energy such as
\begin{equation}
\sigma^2 \propto  \frac{M_\mathrm{dyn}}{R} \propto \frac{M_\mathrm{dyn}}{L}IR, 
\label{eq:bal}
\end{equation}
 where $\sigma$ is the velocity dispersion of a galaxy, $R$ is the galaxy size, $M_\mathrm{dyn}/L$ is the dynamical mass-to-light ratio, and $I$ is the surface brightness ($I\propto L/R^2$). According to this condition, ETGs form a plane, known as the fundamental plane \citep[FP;][]{Djorgovski1987,Dressler1987}, in the space of three observational quantities:
\begin{equation}
\log_{10}R_e=a\log_{10}\sigma_0+b\mu_e+c,
\label{eq:fp}
\end{equation}
in which $R_e$ is the half-light radius, $\sigma_0$ is the central velocity dispersion, and $\mu_e$ is the mean surface brightness within $R_e$ ($\mu_e=-2.5\log_{10}I_e$). 

The FP forms a basis of scaling relations of ETGs, since many other scaling relations on ETGs are variations or projections of the FP. For example, two-dimensional scaling relations on ETGs that were found earlier, such as the Faber--Jackson relation\footnote{A relation between the luminosity and the velocity dispersion} \citep{Faber1976} and the Kormendy relation\footnote{A relation between the size and the surface brightness} \citep{Kormendy1977}, are now regarded as projections of the FP. The mass--size (or luminosity--size) relation of ETGs \citep{Shen2003,Yoon2017} is also a scaling relation related to the FP. 

 If all ETGs are fully virialized and perfectly homologous with a constant $M_\mathrm{dyn}/L$, the coefficients $a$ and $b$ of the FP should be $2$ and $0.4$, respectively, according to the virial relation (Equation \ref{eq:bal}). However, previous studies based on observational data found smaller values of $a$ ($\sim1$--$1.5$) and $b$ ($\sim0.3$) \citep{Bernardi2003a,Jun2008,Hyde2009,LaBarbera2010,Cappellari2013,Saulder2013}. The discrepancy between observational values and expectations from the virial relation is named the tilt of the FP. Determining the origin of the tilt of the FP is directly related to understanding the nonhomology of ETGs and how ETGs are formed and evolved \citep{Bertin2002,Trujillo2004,Cappellari2006,Dekel2006,Robertson2006,Hopkins2008a}. For instance, different degrees of dissipation effects in the formation of ETGs of different masses can cause $M_\mathrm{dyn}/L$ dependence as a function of mass, thereby making the tilt of the FP \citep{Dekel2006,Robertson2006,Hopkins2008a}. 

 FP can be used as a standard ruler to measure distances to ETGs \citep{Gavazzi1999,Mutabazi2014,Im2017}, since the two parameters on the right side of the FP (Equation \ref{eq:fp}) can be derived without distance information. Using the distances of ETGs calculated from the FP, it is possible to derive a cosmological parameter such as the Hubble constant \citep{Hjorth1997,Kelson2000}.

The tightness of FP is an interesting feature of the FP. The FP has a scatter of $\sim20\%$ ($\sim0.08$ -- $0.09$ dex) in the direction of the galaxy size \citep{Bernardi2003a,Hyde2009,LaBarbera2010}, which is also directly translated into the accuracy of distance measurements. Several previous studies investigated the residual of the FP in the sense of stellar populations and found the correlation between the age (or $\alpha$ abundance) and the residuals of the FP \citep{Prugniel1996,Forbes1998,Terlevich2002,Gargiulo2009,Graves2009a,Falcon-Barroso2011,Magoulas2012}. However, the previous studies have not focused yet on how the size of the scatter in the FP depends on properties of ETGs, despite that the investigation of the size of the scatter and its relation to properties of ETGs can improve the standard ruler for cosmic distances by finding the subpopulation of ETGs having small scatters or can give useful information on the formation histories of ETGs.

Therefore, unlike previous studies, we examine the scatter in the FP of ETGs with different properties such as age and structure. By doing so, we find subpopulations of ETGs having smaller (or larger) scatters in the FP. Finally, we discuss the possible origin of compact young ETGs with excessive light for a given mass that have a large scatter in the FP.

 Throughout this study, we use \emph{H$_0=70$} km s$^{-1}$ Mpc$^{-1}$, $\Omega_{\Lambda}=0.7$, and $\Omega_\mathrm{m}=0.3$ as cosmological parameters and the AB magnitude system. 
\\

\section{Sample}\label{sec:sample}
We used the galaxies having spectroscopic redshifts from the Sloan Digital Sky Survey (SDSS) and classified as ETGs in the KIAS value-added catalog \citep{Choi2010}, which is based on SDSS  Data Release 7 \citep[DR7;][]{Abazajian2009}. The catalog classified galaxies as early type or late type based on $u - r$ color, $g-i$ color gradient, and inverse concentration index ($C_\mathrm{inv}$) in the $i$ band \citep[see][]{P&C2005}. Specifically, ETGs are classified by the following criteria: (1) low value of $C_\mathrm{inv}$, because the light is centrally concentrated for ETGs in general; and (2) slightly negative $g-i$ color gradient (bluer outside) and red $u-r$ as is the case for most ETGs, or blue $u-r$ and positive $g-i$ color gradient to allow the inclusion of blue ETGs. The completeness and reliability of the classification is $\sim90\%$. To improve this classification further, the visual inspection was also performed, correcting the morphology of $7\%$ of the inspected galaxies \citep{Choi2010}. We note that our main results are not sensitive to the definition of ETGs, since use of different classification criteria for ETGs  \citep[e.g., criteria in][]{Saulder2013} does not change our results. 

For the final ETG sample\footnote{A total of 16,283 ETGs} used in this study, we checked weight values (from 0 to 1) of the de Vaucouleurs fit component in the combined model of de Vaucouleurs + exponential disk fit for $r$ band ($fracDev\_r$ from SDSS). By doing so, we found that $70\%$ of ETGs have $fracDev\_r>0.95$ and $94\%$ of them have $fracDev\_r>0.71$.\footnote{In the case of $z$ band, $70\%$ of ETGs have $fracDev\_r>0.99$ and $94\%$ of them have $fracDev\_r>0.77$.} So, our ETGs are essentially bulge-dominated galaxies whose surface brightness profiles are well described by the de Vaucouleurs model. 

The spectroscopic redshift range used in this study is $0.025\le z_\mathrm{spec}<0.055$. We set the lower limit of 0.025, which corresponds to the distance of $\sim100\,\mathrm{Mpc}$, in order to mitigate the peculiar velocity effects that can distort distance-dependent galaxy properties at very low redshifts. 

The absolute magnitudes were calculated according to 
\begin{equation}
M=m -DM-K+Qz_\mathrm{spec}, 
\label{eq:abm}
\end{equation}
in which $m$ is the galactic-extinction-corrected apparent magnitude, $DM$ indicates the distance modulus, and $K$ is the $k$-correction. $Qz_\mathrm{spec}$ is a correction term for the passive evolution of galaxy luminosity, where $Q$ is the evolutionary parameter. Model magnitudes from the de Vaucouleurs fits are used for $m$. The galactic extinctions were corrected based on the dust maps of \citet{Schlegel1998}. The $k$-correction values were derived using an IDL software of \citet{Blanton2007}. This software calculates $k$-correction values by fitting spectral energy distribution (SED) models of various ages and metallicities to magnitudes of the five bands of SDSS. The SED models are based on \citet{Bruzual2003} stellar population models\footnote{\url{http://www.bruzual.org/bc03/}} with a \citet{Chabrier2003} initial mass function (IMF). 

The correction for passive evolution of galaxy luminosity is negligible in this study, since we used ETGs at very low redshifts. Thus, we determined $Q$ in a very simple way: using simple stellar populations (SSPs) of \citet{Bruzual2003} models with a \citet{Chabrier2003} IMF. We calculated the changes  in luminosities during 2.4 Gyr owing to the passive evolution in SSPs with ages of $10$--$13$ Gyrs and metallicities of $Z=0.008$--$0.05$. By doing so, we found that, on average, $Q$ values of 1.26, 1.13, 1.07, and 1.02 correspond to the passive evolutions in magnitudes of $g$, $r$, $i$, and $z$ bands, respectively. We note that these values are slightly larger than those used in \citet{Bernardi2003a}. The same $k$- and evolution corrections are applied in $g-r$ color values used here. In this study, we used ETGs with $M_r\le-19.5$, which corresponds to $m_r\approx17.77$ (the magnitude limit for the spectroscopic target selection) at the upper redshift limit of $z_\mathrm{spec}=0.055$. The total number of ETGs in the volume-limited sample of $M_r\le-19.5$ and $0.025\le z_\mathrm{spec}<0.055$ is 22,474.

The physical sizes (in kpc) of galaxies used here are half-light radii (or effective radii) of the de Vaucouleurs fits from SDSS, which are calculated by
\begin{equation}
R_e=a_\mathrm{deV}\sqrt{b/a},
\label{eq:size}
\end{equation}
where $a_\mathrm{deV}$ and $b/a$ are the semi-major axis length and the axis ratio from the de Vaucouleurs fit, respectively. In this study, we only used galaxies with $b/a\footnote{In $r$ band}\ge0.3$ to exclude edge-on galaxies. By this $b/a$ cut, 1196 galaxies are excluded, so that the number of ETGs is 21,278.

We used stellar velocity dispersions that are aperture corrected to $r_{\mathrm{ang}}/8$ by the correction equation in \citet{Jorgensen1995}:  
\begin{equation}
\sigma_0=\sigma_\mathrm{est}\left(\frac{r_\mathrm{fiber}}{r_\mathrm{ang}/8}\right)^{0.04}, 
\label{eq:vel}
\end{equation}
where $\sigma_\mathrm{est}$ is the estimated velocity dispersion in SDSS, $r_{\mathrm{ang}}$ is the angular half-light radius in arcseconds, and $r_\mathrm{fiber}$ is the radius of fibers ($r_\mathrm{fiber}=1.5\arcsec$). 

The instrumental dispersion (spectroscopic sampling) of the SDSS spectrograph is 69 km s$^{-1}$ pixel$^{-1}$, and the resolution of SDSS galaxy spectra measured from the stellar template spectra is $\sim90$ km s$^{-1}$ \citep{Bernardi2003b}. Moreover, some previous studies show that low velocity dispersions less than $\sim100$ km s$^{-1}$ are not reliable \citep{Bernardi2003b,Saulder2013}. Therefore, we only used ETGs with $100\le\sigma_0<420$ km s$^{-1}$. The upper limit is set to be $420$ km s$^{-1}$, since SDSS used template spectra convolved to a maximum velocity dispersion of $420$ km s$^{-1}$ for the velocity dispersion measurements, so that use of velocity dispersions larger than $420$ km s$^{-1}$ is not recommended. We note that an adjustment of the lower limit from $100$ to $70$ km s$^{-1}$ does not essentially change our main results. Applying the $\sigma_0$ cut, the number of ETGs is 16,793.

$\mu_e$ is derived by 
\begin{equation}
\mu_e=m^c  + 2.5\log_{10}(2\pi r_\mathrm{ang}^2) -2.5\log_{10}(1+z_\mathrm{spec})^3,
\label{eq:mu}
\end{equation}
in which $m^c =m - K + Qz_\mathrm{spec}$, while the last one is a correction term for the cosmological dimming of surface brightness in the AB magnitude system.

We note that the magnitude in each band, $a_\mathrm{deV}$, $b/a$, and $\sigma_\mathrm{est}$ are from the catalogs (PhotObjAll and SpecObjAll) of DR15 \citep{Aguado2019}.

 \begin{figure}
\includegraphics[width=\linewidth]{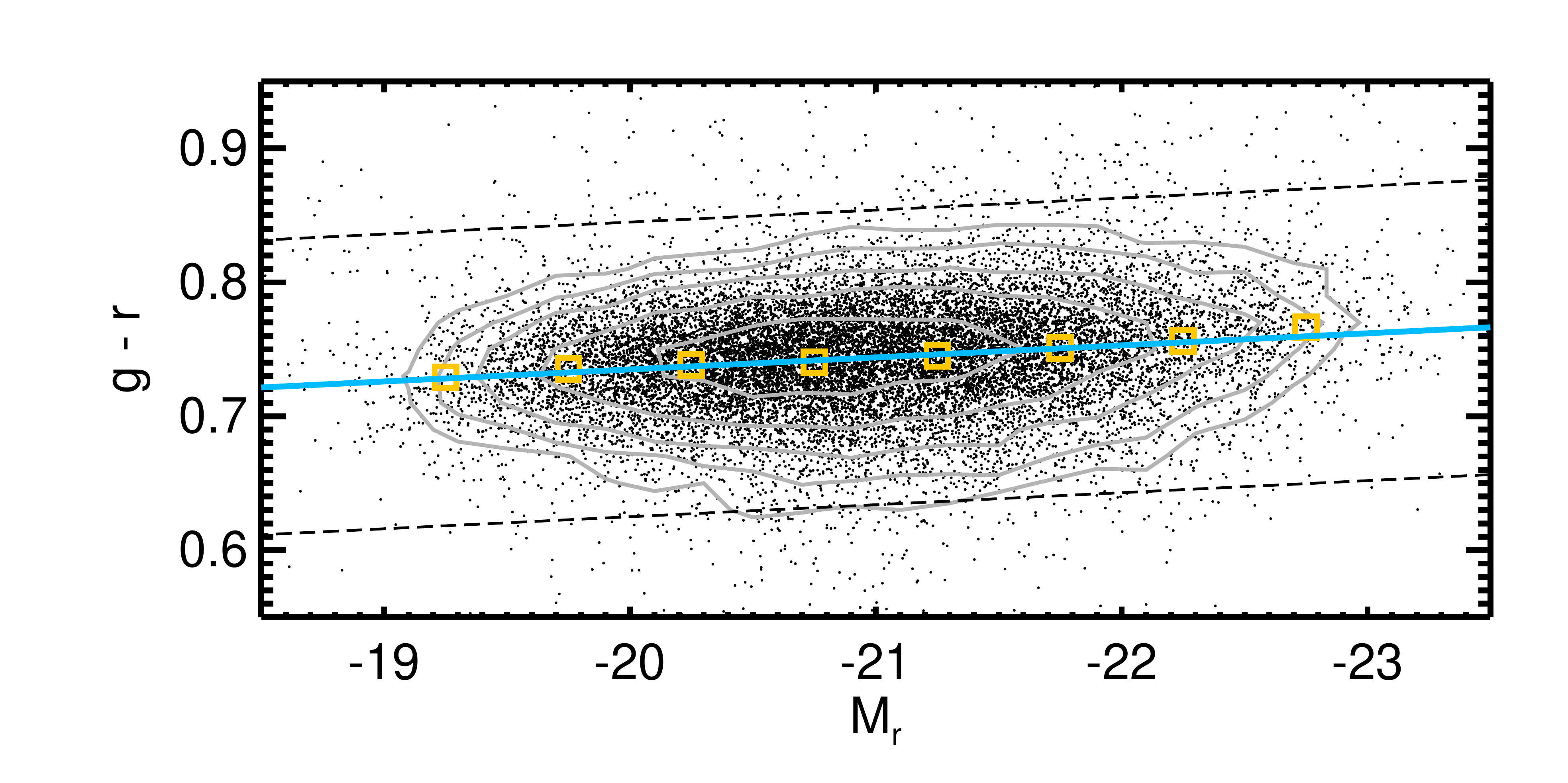}  
\centering
\caption{CMD ($g-r$ vs. $M_r$) for ETG samples used in this study. ETGs with $M_r>-19.5$ are included in this figure for illustrative purposes. The black dots are the ETG samples, while the yellow squares indicate average $g-r$ values in each magnitude bin after $3\sigma$ clipping. The blue solid line denotes a line derived by the iterative least-squares fit with clipping outliers. The dashed lines are upper and lower $3\sigma$ lines ($\sigma=0.0367$), in which $\sigma$ is the standard deviation from the fitted line. The levels of the contour (the gray lines) represent 16, 32, 64, 128, and 256 galaxies in two-dimensional bins of 0.2 (in $M_r$) $\times$ 0.02 (in $g-r$).
\label{fig:cmd}}
\end{figure} 

Figure \ref{fig:cmd} shows the color-magnitude diagram (CMD; $g-r$ vs. $M_r$) for the ETG sample used in this study. Our ETGs form a tight red sequence showing that $g-r$ color values are clustered within $\sim0.2$. We fitted a linear line (red sequence) to the ETGs using the least-squares method. The fitting was conducted iteratively excluding $3\sigma$ outliers from the line, until the number of galaxies converges. The equation of the fitted line (the blue line in Figure \ref{fig:cmd}) is
\begin{equation}
(g-r)=0.555-0.00899M_r.
\label{eq:cmd}
\end{equation}
The derived standard deviation of $g-r$ from the line is 0.0367, so that most of the ETGs are within $\pm0.110$ from the line (see the dashed lines in Figure \ref{fig:cmd}). In this study, we used ETGs whose $g-r$ color values do not deviate more than $3\sigma$ (0.110) from the line. In doing so, all ETGs are contained in the tight red sequence, which means that they are quite uniform in the optical color. Excluding a small number (510) of galaxies by this cut, the total number of ETGs in the final sample is 16,283.

For the galaxy structure parameter that defines subpopulations of ETGs, we use $C_\mathrm{inv}$ from the KIAS value-added catalog. $C_\mathrm{inv}$ is defined as $R_{p, 50}/R_{p, 90}$, in which $R_{p, 50}$ and $R_{p, 90}$ are the seeing-corrected radii at $i$ band containing $50\%$ and $90\%$ of the Petrosian flux, respectively. Thus, a smaller $C_\mathrm{inv}$ means a more compact galaxy structure (or light distribution). The distribution of $C_\mathrm{inv}$ of ETGs shows a peak at $\sim0.31$ and is skewed to larger values, having a median of $0.33$ \citep{P&C2005,Bailin2008}. \citet{P&C2005} adopted $C_\mathrm{inv}<0.43$ for one of the criteria to select ETGs. $99\%$ of the ETGs used here have $C_\mathrm{inv}$ values less than 0.43. 

 \begin{figure}
\includegraphics[width=\linewidth]{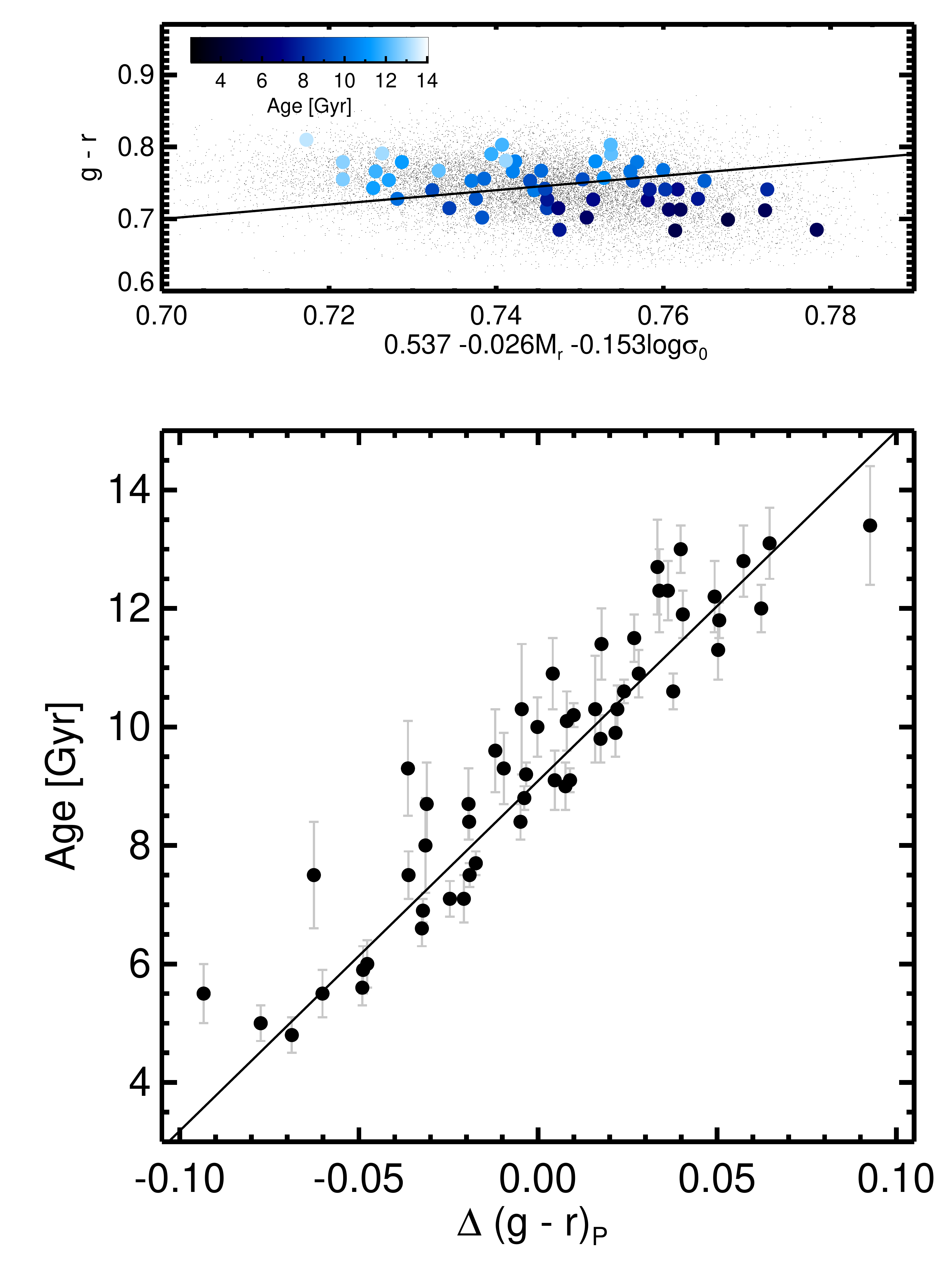}  
\centering
\caption{Top: edge-on view of a plane of constant age (indicated by the solid line; Equation \ref{eq:plane}), together with the distributions of our ETGs (black dots) and the groups of quiescent galaxies (blue circles) from \citet{Graves2009b}. The stellar population age is color-coded (see the color scale for age). In the three-dimensional parameter space, age varies rapidly with the deviation of $g-r$ from the plane ($\Delta (g-r)_\mathrm{P}$). Bottom: age as a function of $\Delta (g-r)_\mathrm{P}$ for the groups of galaxies (black circles) from \citet{Graves2009b}. The gray vertical bars denote errors in age. The black line was fitted with the $\chi^2$ minimization method. 
\label{fig:age}}
\end{figure} 

Previous studies show that a color residual for a given luminosity in CMD is well correlated with the stellar population age for galaxies in red sequence \citep{Cool2006,Connor2019}. Moreover, \citet{Graves2009b} discovered that for quiescent galaxies $g-r$ has the best correlation with age, while the stellar velocity dispersion has a weaker correlation with age than $g-r$.\footnote{See also Figure 22 in \citet{Cappellari2016} for several relations between galaxy stellar populations and galaxy structures.} Motivated from these studies, we find the best proxy for stellar population age of ETGs using three parameters: $g-r$,  $M_r$, and $\log\sigma_0$. For this purpose, we use stacked galaxy properties from \citet{Graves2009b}. \citet{Graves2009b} grouped quiescent galaxies at $0.04<z<0.08$ in SDSS data into 54 bins according to color, velocity dispersions, and magnitudes. They stacked all the spectra of the galaxies in each group to make very high signal-to-noise ratio ($\sim100$--$900$) spectra, which is essential for deriving accurate stellar population properties \citep{Cardiel1998}. So, each group contains up to $\sim1000$ galaxies. Then, they derived luminosity-weighted stellar population age (and metallicity) from the absorption lines of the stacked spectra. 

Using these galaxy groups in the three-dimensional parameter space ($g-r$,  $M_r$, and $\log\sigma_0$), we find a plane from which deviation in the direction of $g-r$ has the highest correlation with age. This plane was identified by finding a combination of coefficients of the plane equation (e.g., Equation \ref{eq:plane}) that yields the minimum $\chi^2$ in the line fitting conducted on the groups of quiescent galaxies in the age versus the deviation of $g-r$ from the plane (see the bottom panel of Figure \ref{fig:age}).

The equation of the plane that corresponds to a constant age of 9.1 Gyr, for example, is 
\begin{equation}
(g-r)=0.537-0.026M_r-0.153\log\sigma_0.
\label{eq:plane}
\end{equation}
We note that $g-r$ is the most dominant factor for determining the age of ETGs as found by \citet{Graves2009b}. The top panel of Figure \ref{fig:age} shows an edge-on view of this plane, together with the distribution of our ETGs (black dots) and the groups of red sequence (quiescent) galaxies from \citet{Graves2009b} (blue circles). This panel shows that the age of ETGs varies rapidly with the deviation of $g-r$ from the constant age plane (hereafter $\Delta (g-r)_\mathrm{P}$). The bottom panel of Figure \ref{fig:age} shows ages as a function of $\Delta (g-r)_\mathrm{P}$ for the groups of galaxies from \citet{Graves2009b}. We find that the standard deviation of the age from the fitted line is only $0.8$ Gyr and the linear Pearson correlation coefficient of the two parameters is 0.93, which means that the stellar population age is well correlated with $\Delta (g-r)_\mathrm{P}$. Therefore, we use $\Delta (g-r)_\mathrm{P}$ as an indicator for the stellar population age of ETGs.\footnote{Age measurements can be largely dependent on which model is used. So, what is more important is relative differences, not the absolute values of ages. Therefore, we use $\Delta (g-r)_\mathrm{P}$ values, rather than using ages directly converted from $\Delta (g-r)_\mathrm{P}$.} We note that the deviation of $g-r$ from the red sequence line (Equation \ref{eq:cmd}) in the CMD in Figure \ref{fig:cmd} is also a good indicator for the stellar population age of ETGs, but not as good as $\Delta (g-r)_\mathrm{P}$ defined in the three-dimensional parameter space. 

\begin{figure*}
\includegraphics[scale=0.45]{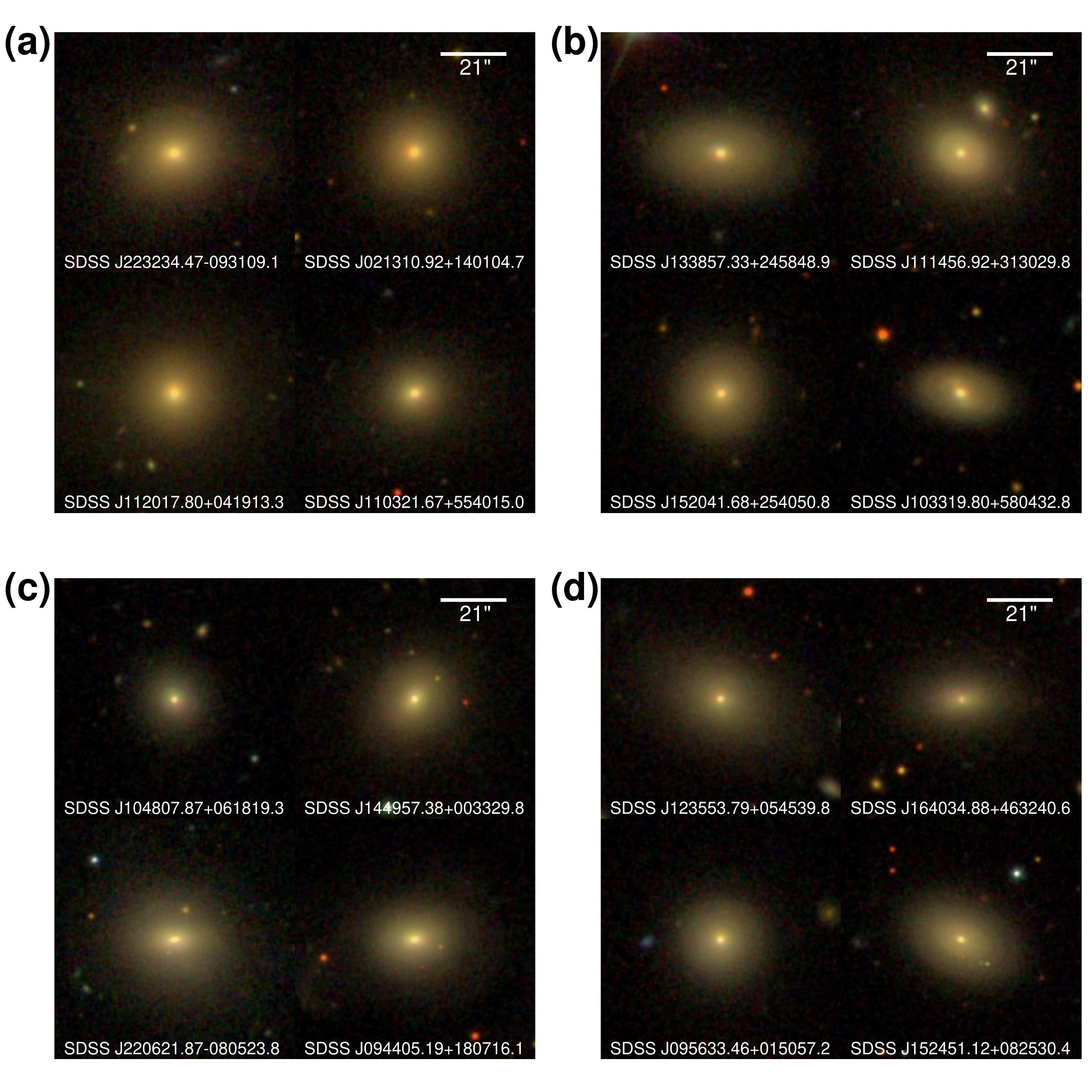}  
\centering
\caption{Color images of ETGs used in this study. (a) ETGs with $\Delta (g-r)_\mathrm{P}>0$ and $C_\mathrm{inv}<0.33$. (b) ETGs with $\Delta (g-r)_\mathrm{P}>0$ and $C_\mathrm{inv}>0.33$. (c) ETGs with $\Delta (g-r)_\mathrm{P}<0$ and $C_\mathrm{inv}<0.33$. (d) ETGs with $\Delta (g-r)_\mathrm{P}<0$ and $C_\mathrm{inv}>0.33$. The image scales are indicated by the horizontal bars. The names of the ETGs are in each image. 
\label{fig:cimg}}
\end{figure*}

A few images of typical ETGs used in this study are presented in Figure \ref{fig:cimg}, in which the ETGs are grouped according to $\Delta (g-r)_\mathrm{P}$ and $C_\mathrm{inv}$.
\\

\section{Fundamental Plane Fitting}\label{sec:fit}

We fit the FP using the plane-fitting code (LTS\_PLANEFIT\footnote{\url{https://www-astro.physics.ox.ac.uk/~mxc/software/}}) that was used in \citet{Cappellari2013}. This code performs an extremely robust fit of planes to data. The basic scheme for the fitting method of the code is $\chi^2$ minimization after trimming outliers. We briefly describe the method in this section. 
\begin{enumerate}
\item Finding the subset of $h$ data points having the smallest $\chi^2$. Here, $h$ is $(N+p+1)/2$, where $N$ is the number of total points and $p$ is the data dimension (for the plane fit, $p=3$). The $\chi^2$ is described by
\begin{equation}
\chi^2=\sum_{i=1}^{n} \frac{(ax_i+by_i+c-z_i)^2}{(a\Delta x_i)^2+(b\Delta y_i)^2+(\Delta z_i)^2+\varepsilon^2},
\label{eq:chi}
\end{equation}
in which $x_i=\log_{10}\sigma_0$, $y_i=\mu_e$, and $z_i=\log_{10}R_e$. $\Delta$ indicates the error of the quantity. $\varepsilon$ is an intrinsic scatter in the direction of $\log_{10}R_e$.
\item Calculating the standard deviation value ($\sigma$) of the residuals for the $h$ data points. Among all the data points ($N$), those that deviate more than $3\sigma$ from the fitted plane are excluded.
\item Iterating the above steps until the number of the selected data points converges.
\item Computing the $\chi^2$ for the data points.
\item All the above steps are repeated varying $\varepsilon$ until $\chi^2=\nu$, in which $\nu$ is the degree of freedom.
\end{enumerate}
By this process, we determined coefficients of the FP with the intrinsic scatter in the direction of $\log_{10}R_e$. Compared to the standard $\sigma$ clipping method, the fitting result derived by this method is less affected by outliers that deviate significantly from other data, since the process used here clips data points from inside out, which is opposite to the standard $\sigma$ clipping method \citep{Cappellari2013}.
\\

\section{Results}
\subsection{Scatter in the Fundamental Plane}\label{sec:results:scatter}

 \begin{figure}
\includegraphics[width=\linewidth]{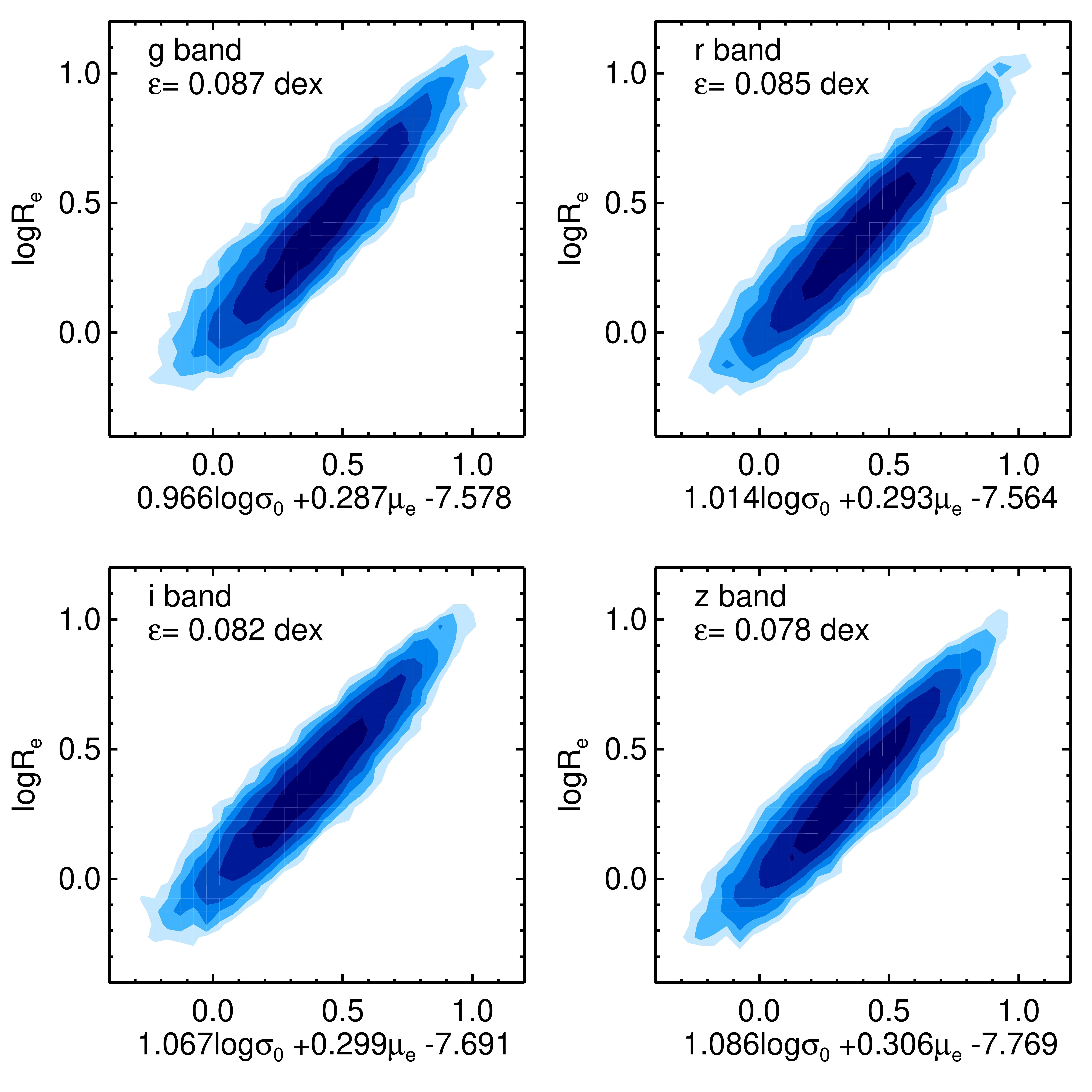}  
\centering
\caption{Edge-on views of the FPs of all ETGs in the four bands ($g$, $r$, $i$, and $z$). The intrinsic scatter values ($\varepsilon$) in the direction of $\log_{10}R_e$ are indicated in each panel. The levels of the contour represent 6, 12, 24, 48, 96, and 192 galaxies in two-dimensional bins of 0.05 (in $x$ axis) $\times$ 0.05 (in $y$ axis).
\label{fig:FPall}}
\end{figure} 

In this section, we present results on the scatter in the FPs for ETGs with various properties. All the values of scatters in the FPs and their coefficients (and their errors) shown here are summarized in Tables \ref{tb:g}, \ref{tb:r}, \ref{tb:i}, and \ref{tb:z} in the Appendix \ref{AppendixTable} for $g$, $r$, $i$, and $z$ bands, respectively.

The edge-on view of the FPs of ETGs in the four bands is shown in Figure \ref{fig:FPall}. The $\varepsilon$ values of the FPs are within $\sim0.08$ -- $0.09$ dex and smaller in the redder band (the smallest in $z$ band), which are consistent with previous results based on large SDSS data \citep{Bernardi2003a,Hyde2009,LaBarbera2010}.

 \begin{figure}
\includegraphics[width=\linewidth]{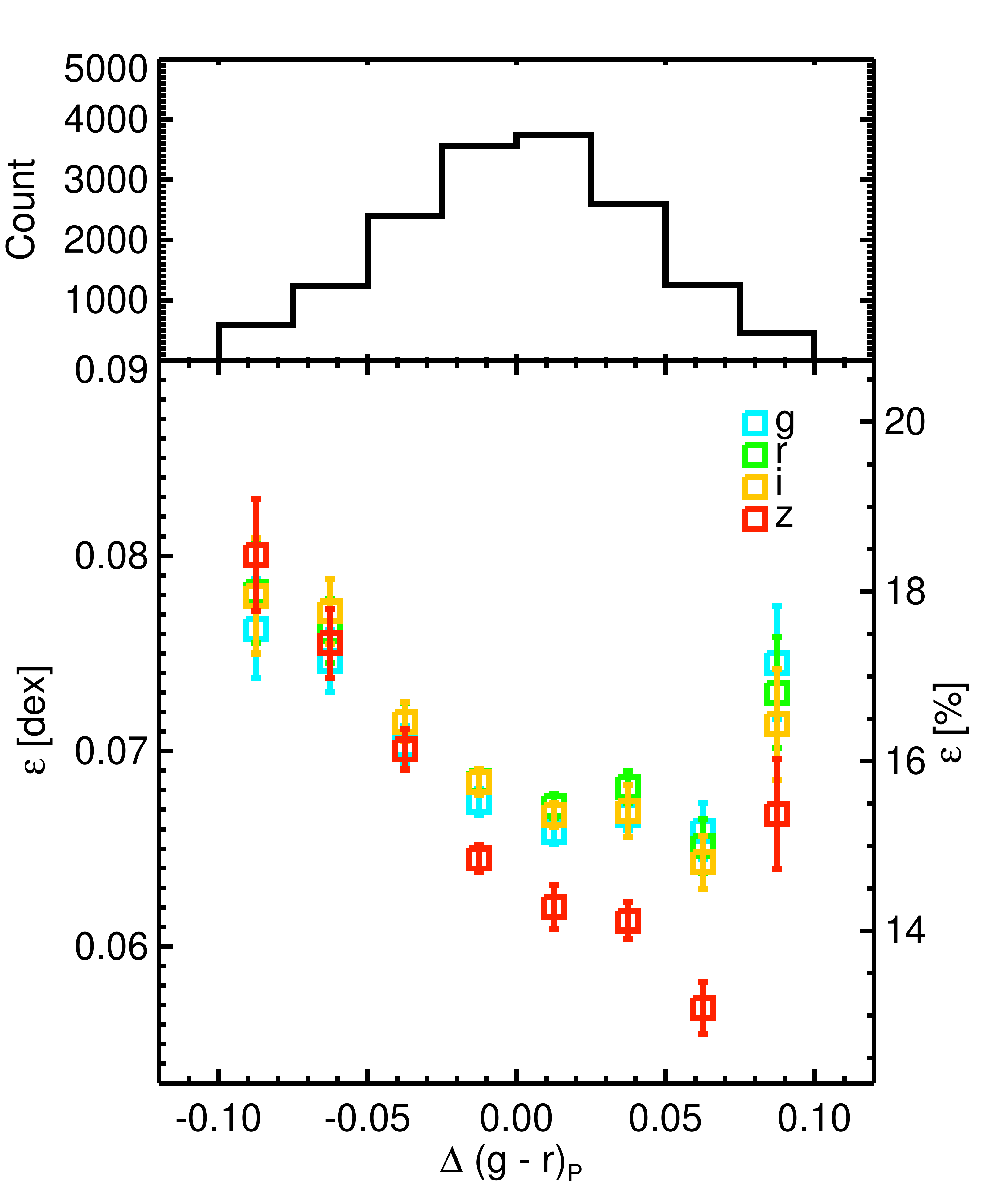}  
\centering
\caption{Intrinsic scatters in the FPs as a function of $\Delta (g-r)_\mathrm{P}$. The histogram in the top panel shows the distribution of $\Delta (g-r)_\mathrm{P}$. The left $y$-axis indicates $\varepsilon$ in the logarithmic scale, which is converted to $\varepsilon$ in the linear scale (in unit of $\%$) in the right $y$-axis.
\label{fig:scolor}}
\end{figure} 
 \begin{figure}
\includegraphics[width=\linewidth]{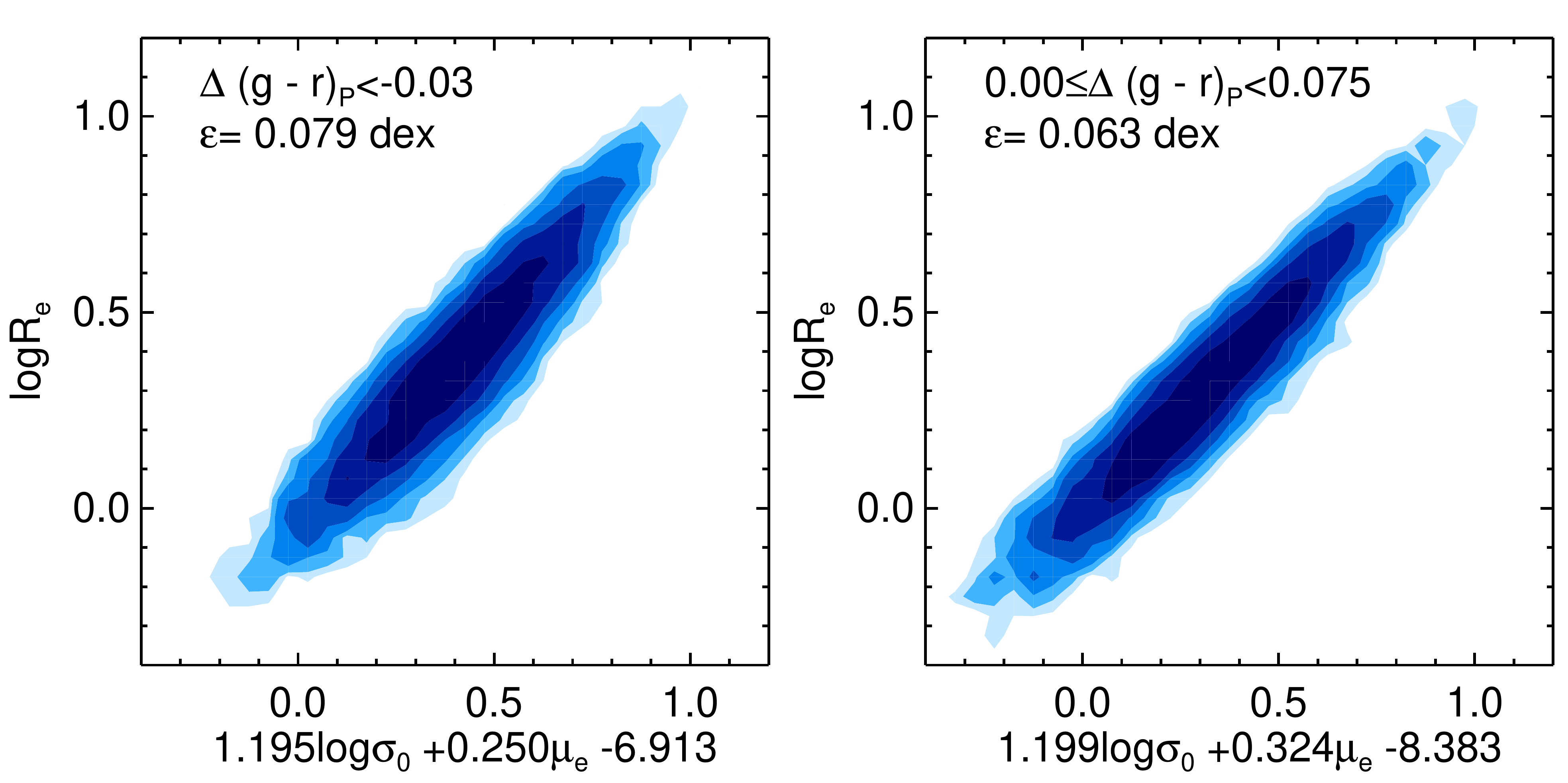}  
\centering
\caption{Edge-on views of $z$-band FPs for ETGs in different $\Delta (g-r)_\mathrm{P}$ bins. $\varepsilon$ values are indicated in each panel. The levels of the contour in each panel are scaled down from the values in Figure \ref{fig:FPall} by a ratio between the number of ETGs in each bin and the number of whole ETGs. 
\label{fig:FPcol}}
\end{figure}

We derived $\varepsilon$ for subpopulations of ETGs divided by $\Delta (g-r)_\mathrm{P}$. Figure \ref{fig:scolor} shows that the scatter in the FP depends strongly on $\Delta (g-r)_\mathrm{P}$ in such a way that ETGs with higher $\Delta (g-r)_\mathrm{P}$ have smaller $\varepsilon$. ETGs with $0.00<\Delta (g-r)_\mathrm{P}<0.08$ have $\varepsilon\lesssim0.07$, while ETGs with $\Delta (g-r)_\mathrm{P}<-0.03$ have $\varepsilon\gtrsim0.07$. The difference in $\varepsilon$ is up to $\sim0.02$ dex ($\sim5\%$) between relatively blue and red ETGs. As in the FPs of whole ETGs, $\varepsilon$ is smaller in redder bands. Figure \ref{fig:FPcol} shows edge-on views of $z$-band FPs for ETGs in different $\Delta (g-r)_\mathrm{P}$ bins.

ETGs in all $\Delta (g-r)_\mathrm{P}$ bins\footnote{Only except the lowest $\Delta (g-r)_\mathrm{P}$ bin in $z$ band} have smaller $\varepsilon$ than whole ETGs (see $\varepsilon$ in Figure \ref{fig:FPall}), which means that dividing ETGs by ages makes scatters in the FP small. This is a reflection of the fact that the residuals of the FP\footnote{The left-hand term of Equation \ref{eq:fp} subtracted by the right-hand term of the same equation.} have a correlation with ages ($\Delta (g-r)_\mathrm{P}$) in the sense that ETGs in the higher FP residuals have younger ages (lower $\Delta (g-r)_\mathrm{P}$) as found in previous studies \citep{Graves2009a,Magoulas2012}. Thus, selecting ETGs with similar ages is roughly identical to selecting ETGs with similar residual values in the FP of whole ETGs (hence the smaller $\varepsilon$).

Although the number is small, ETGs with $0.075\le\Delta (g-r)_\mathrm{P}<0.100$ have a high $\varepsilon$ ($\sim0.07$ dex). The red color of some galaxies in that bin may be due to heavy internal dust extinctions, which means that their genuine color (internal extinction-corrected color) may not be that red. This can be the reason why they have a large $\varepsilon$ comparable to that of blue ETGs. Indeed, we found that  $16\%$ of the ETGs with $0.075\le\Delta (g-r)_\mathrm{P}<0.100$ have signs of dust lanes, and this fraction is higher than the fraction of ETGs that have dust lanes, $\sim4$--$7\%$, which was found based on SDSS data \citep{Kaviraj2010,Kaviraj2012}.

 \begin{figure*}
\includegraphics[scale=0.245]{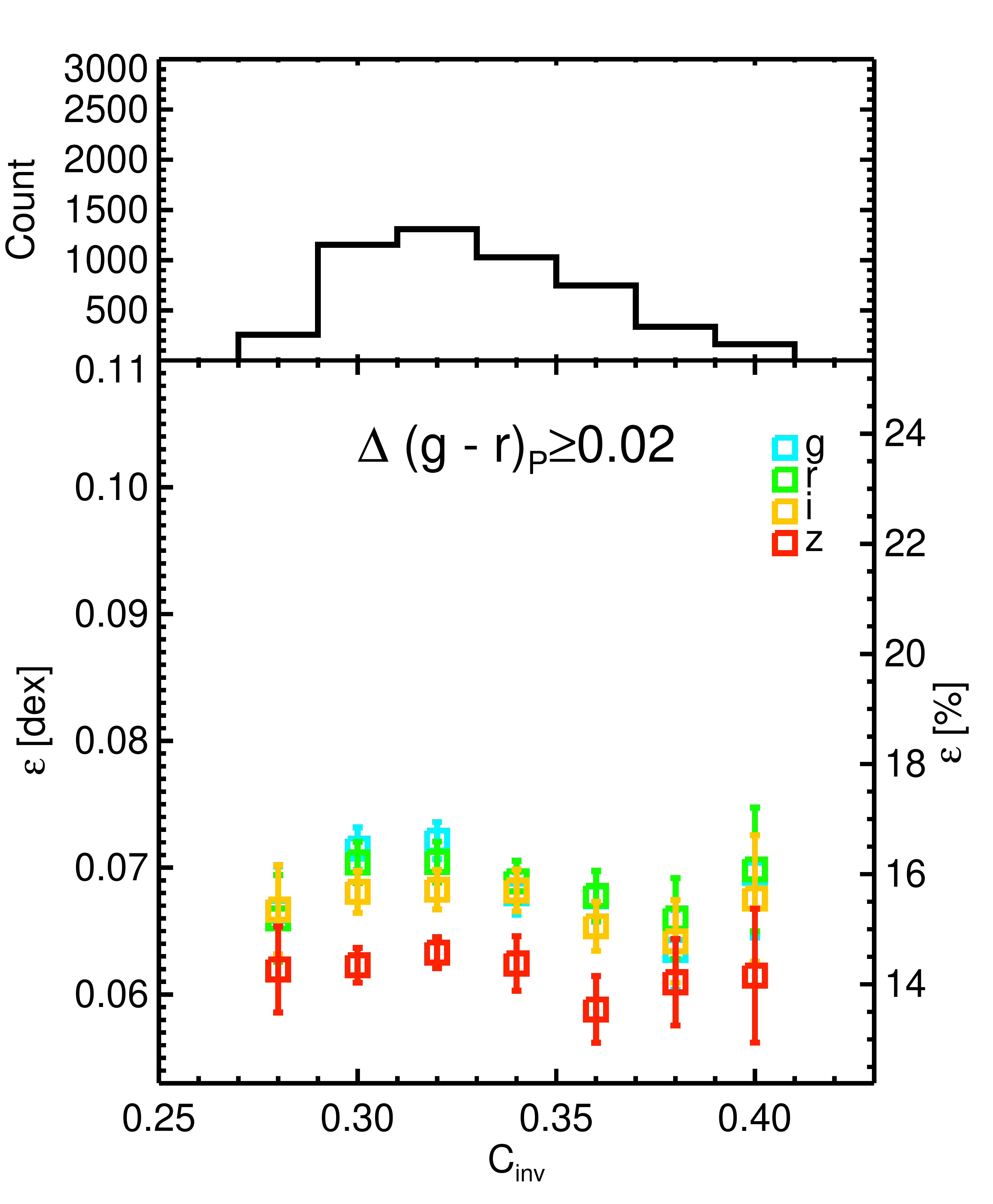}\includegraphics[scale=0.245]{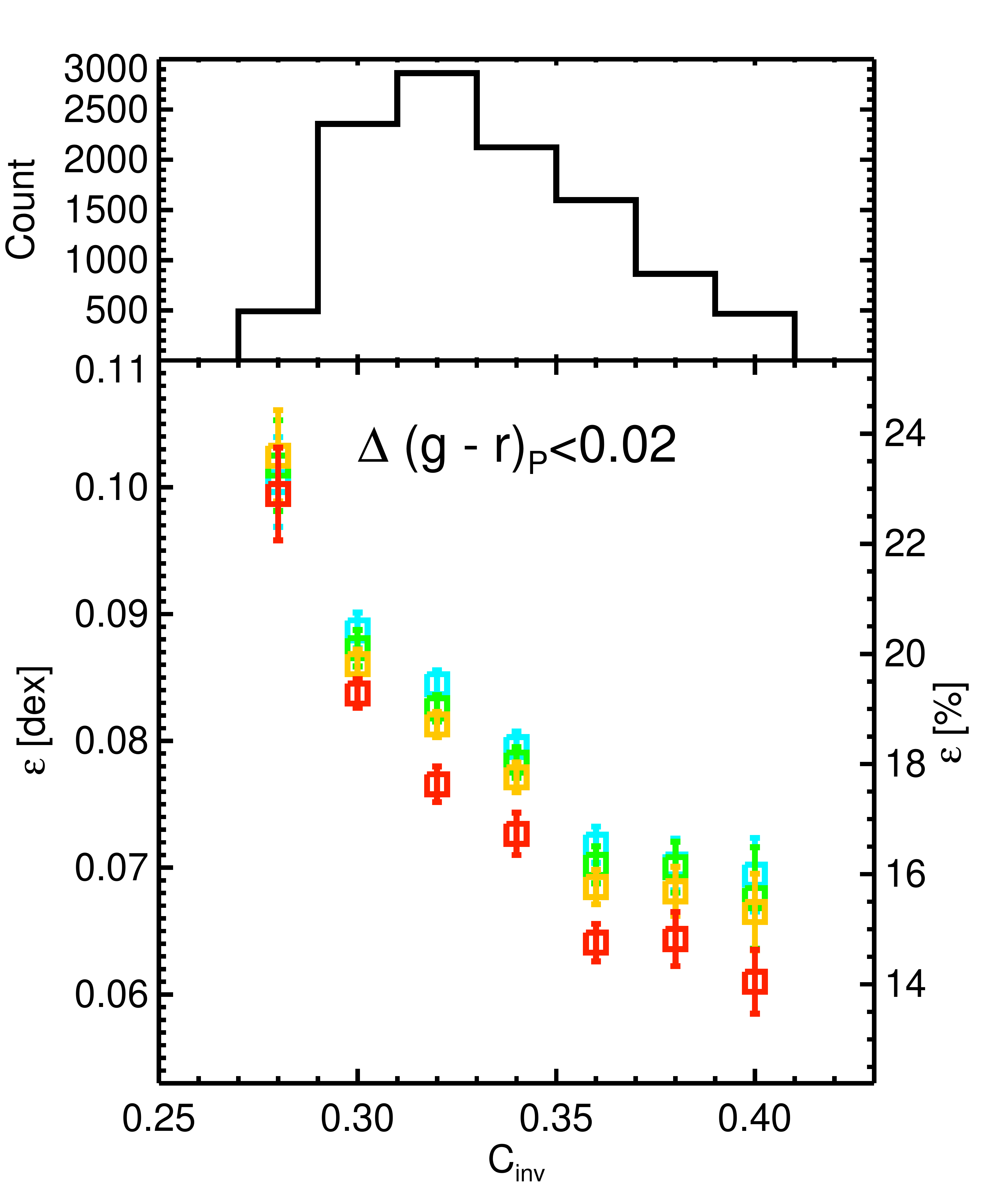}  
\centering
\caption{Intrinsic scatters in the FPs as a function of $C_\mathrm{inv}$. The left panel is for ETGs with $\Delta (g-r)_\mathrm{P}\ge0.02$, while the right panel is for ETGs with $\Delta (g-r)_\mathrm{P}<0.02$. The histograms in the top panels show the distributions of $C_\mathrm{inv}$. The left $y$-axes indicate $\varepsilon$ in the logarithmic scale, which are converted to $\varepsilon$ in the linear scale (in units of $\%$) in the right $y$-axes.
\label{fig:scon}}
\end{figure*} 
 \begin{figure}
\includegraphics[width=\linewidth]{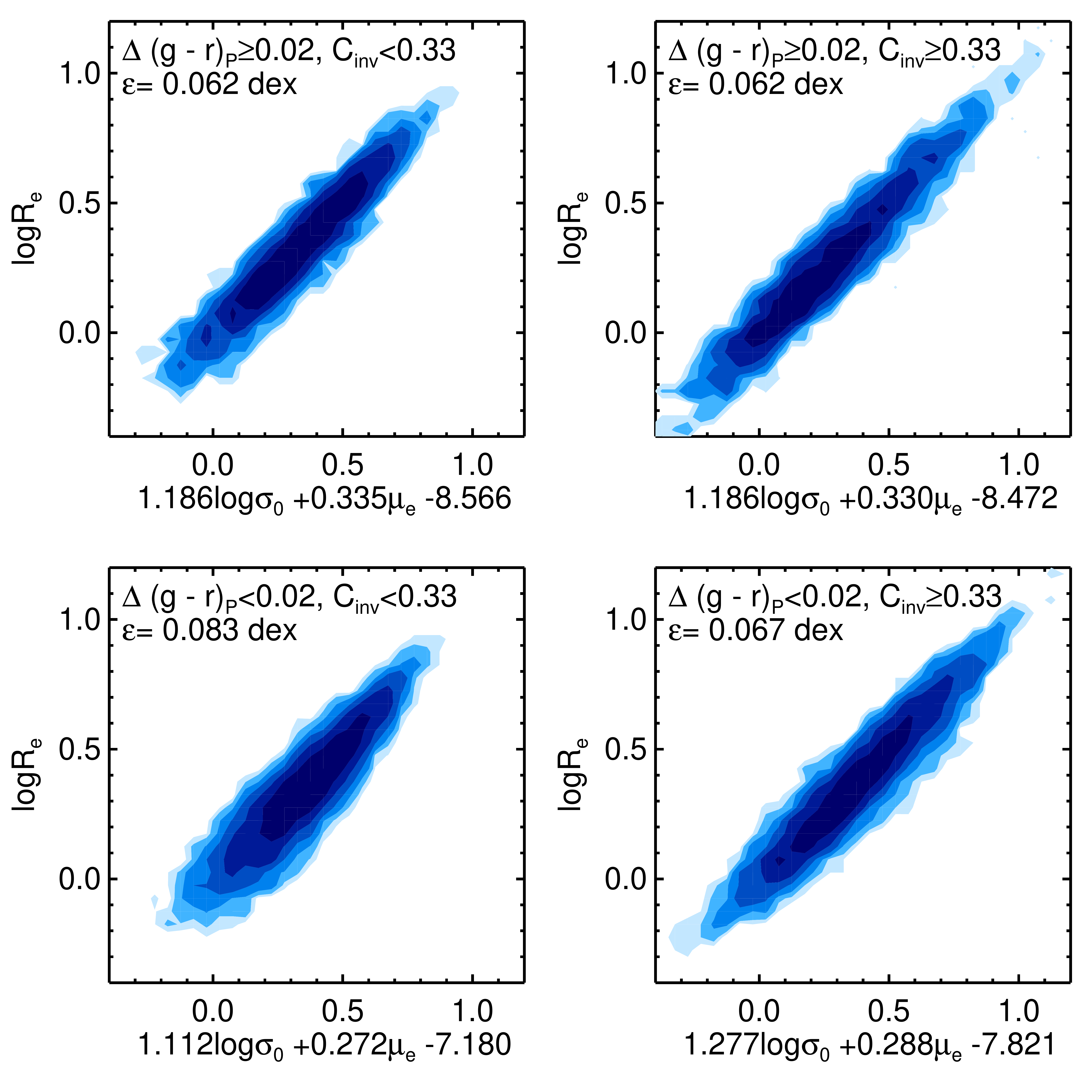}  
\centering
\caption{Edge-on views of $z$-band FPs for ETGs in different $C_\mathrm{inv}$ and $\Delta (g-r)_\mathrm{P}$ bins. $\varepsilon$ values are indicated in each panel. The levels of the contour in each panel are scaled down from the values in Figure \ref{fig:FPall} by a ratio between the number of ETGs in each bin and the number of whole ETGs. 
\label{fig:FPcon}}
\end{figure} 

We further divide ETGs by $C_\mathrm{inv}$. Figure \ref{fig:scon} shows how $\varepsilon$ depends on $C_\mathrm{inv}$ for ETGs with $\Delta (g-r)_\mathrm{P}\ge0.02$ or $\Delta (g-r)_\mathrm{P}<0.02$.\footnote{Here, we divide ETGs at $\Delta (g-r)_\mathrm{P}=0.02$ to display a clear difference between the two groups. We note that use of $\Delta (g-r)_\mathrm{P}=0.00$ for separation gives similar results.} The trends are clearly different in these two groups. $\varepsilon$ of ETGs with $\Delta (g-r)_\mathrm{P}\ge0.02$ are not dependent on $C_\mathrm{inv}$ within $0.01$ dex. On the other hand, in the case of ETGs with $\Delta (g-r)_\mathrm{P}<0.02$, more compact ones have larger $\varepsilon$ than less compact counterparts. The difference in $\varepsilon$ is up to $\sim0.04$ dex ($\sim9\%$) between compact and noncompact ETGs with $\Delta (g-r)_\mathrm{P}<0.02$. In Figure \ref{fig:FPcon}, we present edge-on views of $z$-band FPs for ETGs in different $C_\mathrm{inv}$ and $\Delta (g-r)_\mathrm{P}$ bins. 

Some previous studies argue that coefficients and scatters in the FPs can be affected by different cuts on $M_r$ or $\sigma_0$ \citep{DOnofrio2008,Hyde2009,LaBarbera2010}, which implies that different distributions of $M_r$ or $\sigma_0$ can bias $\varepsilon$ of the FPs. To test how much such a bias from the different distributions of $M_r$ or $\sigma_0$ affects our results, we made the $M_r$ or $\sigma_0$ distributions in all the galaxy categories identical to the distributions of the whole ETG sample and fitted FPs to calculate $\varepsilon$. In the test, we find that $\varepsilon$ values of ETG subpopulations divided by $\Delta (g-r)_\mathrm{P}$ in Figure \ref{fig:scolor} are consistent with the original $\varepsilon$ within $\sim2\sigma$,\footnote{Here, $\sigma$ is the error of $\varepsilon$.} so that the trend shown in Figure \ref{fig:scolor} is still valid. This is also true for the ETG subpopulations in different $C_\mathrm{inv}$ bins in Figure \ref{fig:scon}, except that the maximum difference between $\varepsilon$ values in the left panel of Figure \ref{fig:scon} becomes $\sim0.015$ dex. Therefore, our results on $\varepsilon$ of the FPs summarized in Figures \ref{fig:scolor} and \ref{fig:scon} are not simple by-products of the bias from the different distributions of $M_r$ or $\sigma_0$ in the galaxy categories.

In summary, we find that the scatter in the FP depends on galaxy age in such a way that relatively older ETGs have smaller scatters in the FP than younger ETGs. For young ETGs, less compact ones have smaller scatters in the FPs, whereas for old ETGs, the scatter is independent of the compactness of the light distribution. 
\\

\subsection{Scatter in $M_\mathrm{dyn}/L$}\label{sec:results:ml}

To further understand the scatter in the FPs, we investigate scatters of $M_\mathrm{dyn}/L$ for subpopulations of ETGs, since the scatter in the FP is closely connected with the scatter in $M_\mathrm{dyn}/L$ (see Equations \ref{eq:bal} and \ref{eq:fp} and their description in Section \ref{sec:intro}). Here, we use $i$-band luminosity ($L_i$) in the solar luminosity units. We note that the results based on luminosities in other bands ($g$, $r$, and $z$) are essentially identical. $M_\mathrm{dyn}$ used here is derived by
\begin{equation}
M_\mathrm{dyn}=k\frac{\sigma_0^2R_e}{G}, 
\label{eq:dym}
\end{equation}
in which we use $k=3.8$, since it is known that $k=3.8$ in this equation most accurately traces the true enclosed mass within $R_e$ \citep{Hopkins2008a}. The units of $M_\mathrm{dyn}$ are solar mass.

\begin{figure*}
\includegraphics[scale=0.23]{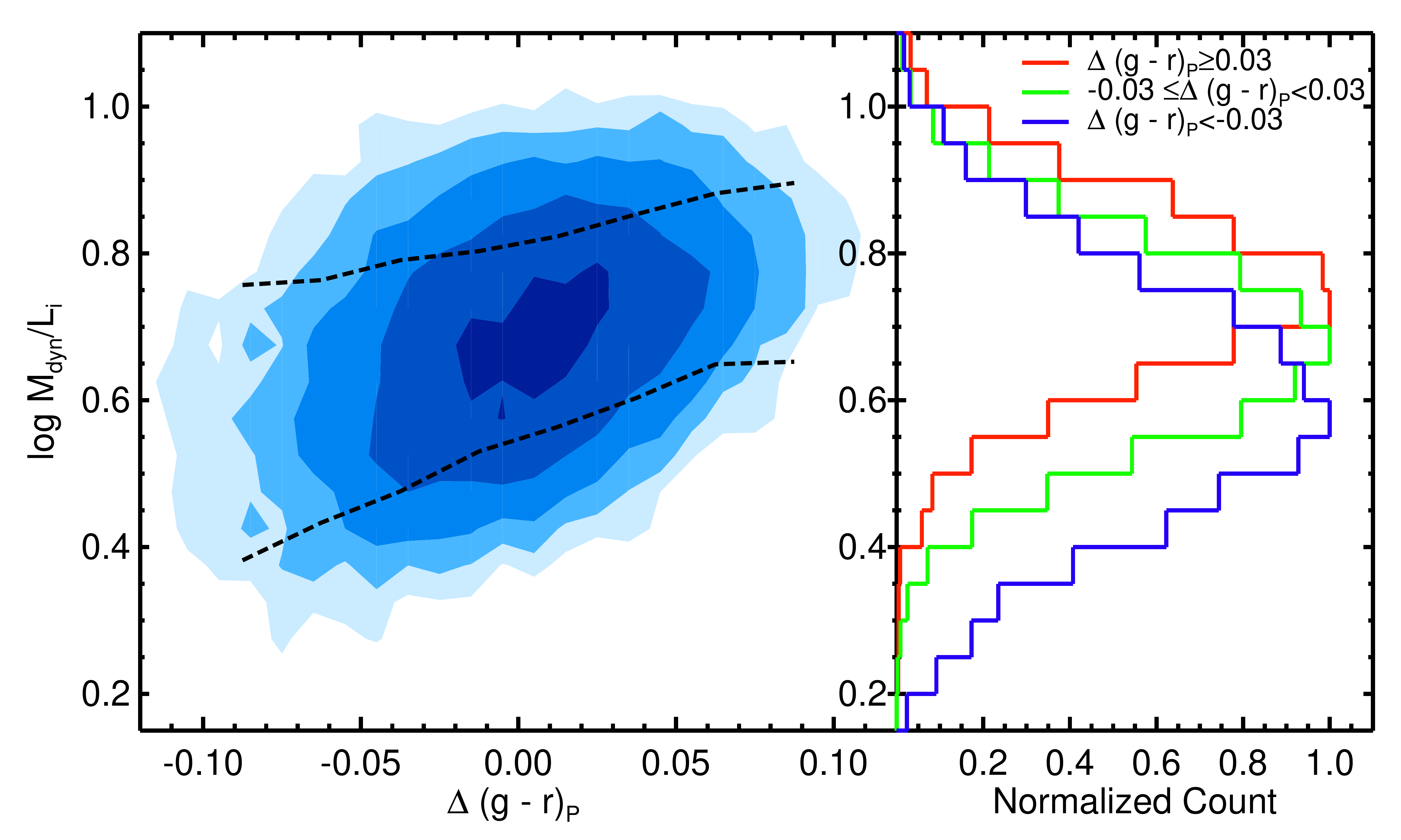}  
\centering
\caption{Two-dimensional distribution of ETGs in the $\log M_\mathrm{dyn}/L_i$ versus $\Delta (g-r)_\mathrm{P}$ plane (left panel) and its projected $M_\mathrm{dyn}/L_i$ distributions for three $\Delta (g-r)_\mathrm{P}$ bins (right panel). The dashed lines in the left panel indicate 84th and 16th percentiles of $\log M_\mathrm{dyn}/L_i$ for a given $\Delta (g-r)_\mathrm{P}$, so that the width between the two dashed lines means the scatter in $\log M_\mathrm{dyn}/L_i$. The levels of the contours represent 6, 12, 24, 48, 96, and 192 galaxies in two-dimensional bins of 0.01 (in $x$-axis) $\times$ 0.05 (in $y$-axis). The peaks of the distributions in the right panel are normalized to 1. 
\label{fig:mlcolor}}
\end{figure*} 

Figure \ref{fig:mlcolor} shows the distribution of ETGs in the $\log M_\mathrm{dyn}/L_i$ versus $\Delta (g-r)_\mathrm{P}$ plane and its projected $\log M_\mathrm{dyn}/L_i$ distributions. The figure shows that ETGs with lower $\Delta (g-r)_\mathrm{P}$ have a larger scatter in $M_\mathrm{dyn}/L_i$, which is consistent with their larger scatter in the FP. Moreover, the $M_\mathrm{dyn}/L_i$ distribution for ETGs with lower $\Delta (g-r)_\mathrm{P}$ is biased to smaller $M_\mathrm{dyn}/L_i$ (up to $\sim0.20$ dex). In other words, young ETGs have a wider range of $M_\mathrm{dyn}/L_i$ (or a larger scatter in the FP) than old ETGs, due to the higher fraction of ETGs having more light for a given dynamical mass (lower $M_\mathrm{dyn}/L_i$).

 \begin{figure}
\includegraphics[scale=0.33]{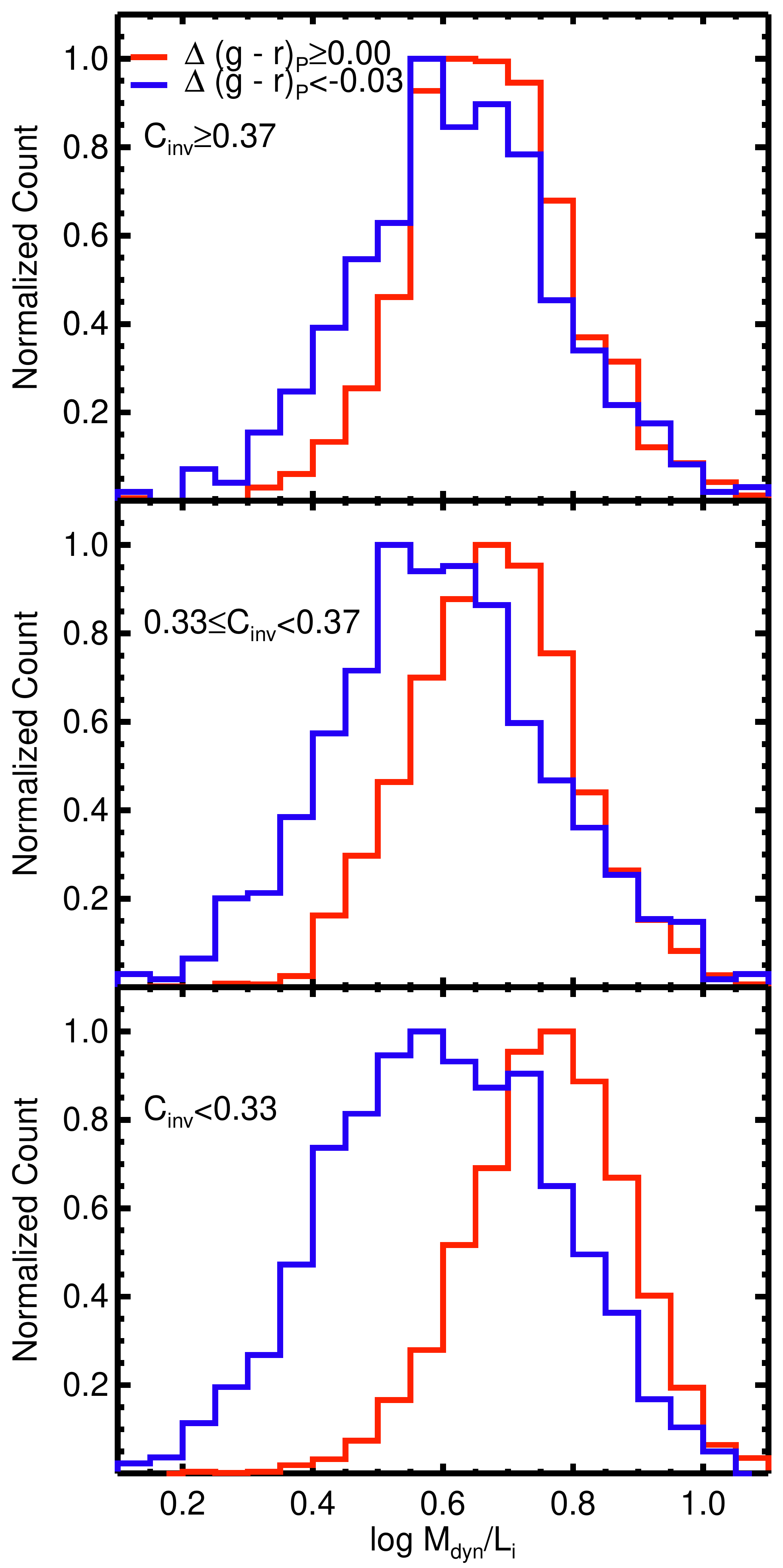}  
\centering
\caption{$M_\mathrm{dyn}/L_i$ distributions of ETG subpopulations divided by $C_\mathrm{inv}$ and $\Delta (g-r)_\mathrm{P}$.  The peaks of the distributions are normalized to 1. 
\label{fig:mlcon}}
\end{figure} 

We also show the $M_\mathrm{dyn}/L_i$ distributions of ETG subpopulations divided by $C_\mathrm{inv}$ in Figure \ref{fig:mlcon}. In the case of noncompact ETGs with $C_\mathrm{inv}\ge0.37$, the $M_\mathrm{dyn}/L_i$ distributions of young ($\Delta (g-r)_\mathrm{P}<-0.03$) and old ($\Delta (g-r)_\mathrm{P}\ge0.00$) ETGs are not much different from each other. However, for compact ETGs with $C_\mathrm{inv}<0.33$, the fraction of ETGs having lower $M_\mathrm{dyn}/L_i$ is significantly higher in young ETGs, so that compact young ETGs have a wider range of $M_\mathrm{dyn}/L_i$ (and a larger scatter in the FP) than the old counterparts. 
\\

\section{Discussion on Compact ETG\lowercase{s}}\label{sec:discussion}

In this study, we find that younger and more compact ETGs have larger scatters in the FP of ETGs than other ETG subpopulations. We also find that compact young ETGs have a wide range of $M_\mathrm{dyn}/L_i$, and this is due to the fact that the fraction of ETGs having more light for a given dynamical mass is high in compact young ETGs. In this section, we focus on the compact ETGs, especially those having excessive light for a given mass, and discuss their origin.  

 \begin{figure}
\includegraphics[width=\linewidth]{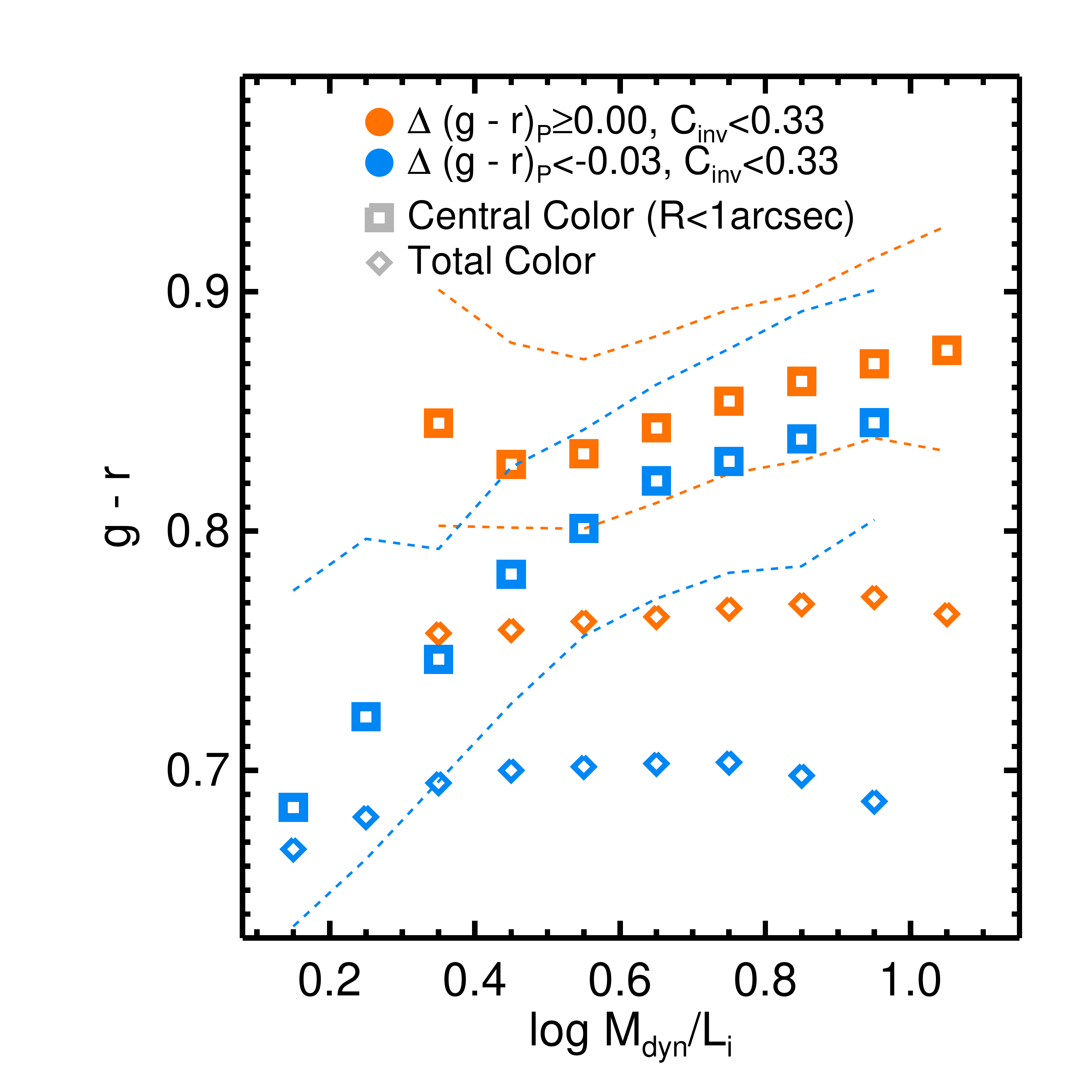}  
\centering
\caption{Median central color (squares) and total color (diamonds) of compact ETGs with $C_\mathrm{inv}<0.33$ as a function of $\log M_\mathrm{dyn}/L_i$. We divide ETGs into two $\Delta (g-r)_\mathrm{P}$ bins. The dashed lines indicate 84th and 16th percentiles of the central color for a given $\log M_\mathrm{dyn}/L_i$. We use fiber magnitudes measured within an aperture of $1\arcsec$ radius to compute central color values.
\label{fig:cml}}
\end{figure} 

To understand more about the property of the compact ETGs with $C_\mathrm{inv}<0.33$, we investigate the central color of compact ETGs as a function of $M_\mathrm{dyn}/L_i$. For the central color, we use fiber magnitudes measured within an aperture of $1\arcsec$ radius. The $1\arcsec$ corresponds to $\sim0.5$--$1$ kpc for our sample. This is a typical size range for compact extra light components that are generally found in ETGs \citep{Hopkins2008b,Hopkins2009,Kormendy2009}. We note that the aperture size is also similar to the size of the point spread function in the SDSS $r$ band.\footnote{The half-width at half-maximum (half of the FWHM) is $0.7\arcsec$.}

In this examination for central color, we excluded the ETGs with active galactic nuclei (AGNs) that can severely affect central color. To do so, we rejected the ETGs whose spectra were classified as AGNs in SDSS based on the optical line ratio diagram of \citet{Baldwin1981}. We found that $\sim4\%$ of compact ETGs with $C_\mathrm{inv}<0.33$ are AGNs. Including or excluding AGNs does not affect the results, since the number of AGNs is small.

Figure \ref{fig:cml} shows the median central color (the squares) and total color (the diamonds) of compact ETGs with $C_\mathrm{inv}<0.33$ as a function of $\log M_\mathrm{dyn}/L_i$. For compact old ETGs with $\Delta (g-r)_\mathrm{P}\ge0.00$, the total color negligibly changes as a function of $M_\mathrm{dyn}/L_i$ within $\sim0.02$. The central color values of compact old ETGs are redder than the total colors, which is a typical color gradient for ETGs \citep{P&C2005,Hopkins2009}. The central color of compact old ETGs is also not substantially dependent on $M_\mathrm{dyn}/L_i$, although it slightly decreases within $\sim0.05$ as $M_\mathrm{dyn}/L_i$ decreases. 

In the case of compact young ETGs with $\Delta (g-r)_\mathrm{P}<-0.03$, the total color value marginally changes as a function of $M_\mathrm{dyn}/L_i$ within $\sim0.04$, which is a similar trend to the total color of compact old ETGs. On the other hand, the central color of compact young ETGs shows a significant change as a function of $M_\mathrm{dyn}/L_i$. For compact young ETGs with $\log M_\mathrm{dyn}/L_i>0.5$, their central color value is not much different from (only $\sim0.02$ bluer than) that of the compact old counterparts. However, the central color of compact young ETGs becomes dramatically bluer at $\log M_\mathrm{dyn}/L_i<0.5$, so that central color is close to the total color at $\log M_\mathrm{dyn}/L_i\sim0.2$. Thus, compact young ETGs with excessive light for a given mass have exceptionally blue central color, while their total color and structure are not that distinctive compared to other ETGs. 

Many previous studies demonstrated that a merger between galaxies with abundant gas makes the gas funnel into the central region of the merger remnant owing to loss of angular momentum via radiation and tidal torques during the merger process, and thereby triggering starburst in the center of the galaxy \citep{Hernquist1989,Barnes1991,Barnes1996,Hopkins2008a}. This process builds central extra light components of young stellar populations with typical sizes of $\sim0.5$--$1$ kpc in the inner regions of post-merger galaxies that make centrally concentrated compact light distributions  \citep{Mihos1994,Robertson2006,Hopkins2008a,Hopkins2008b,Hopkins2009}, central blue color, and low $M_\mathrm{dyn}/L$ \citep{Robertson2006,Hopkins2008a} in the merger remnants. Therefore, compact young ETGs with low $M_\mathrm{dyn}/L$ are likely to have a connection with recent gas-rich major mergers considering their properties examined in this study. 

 \begin{figure*}
\includegraphics[width=\linewidth]{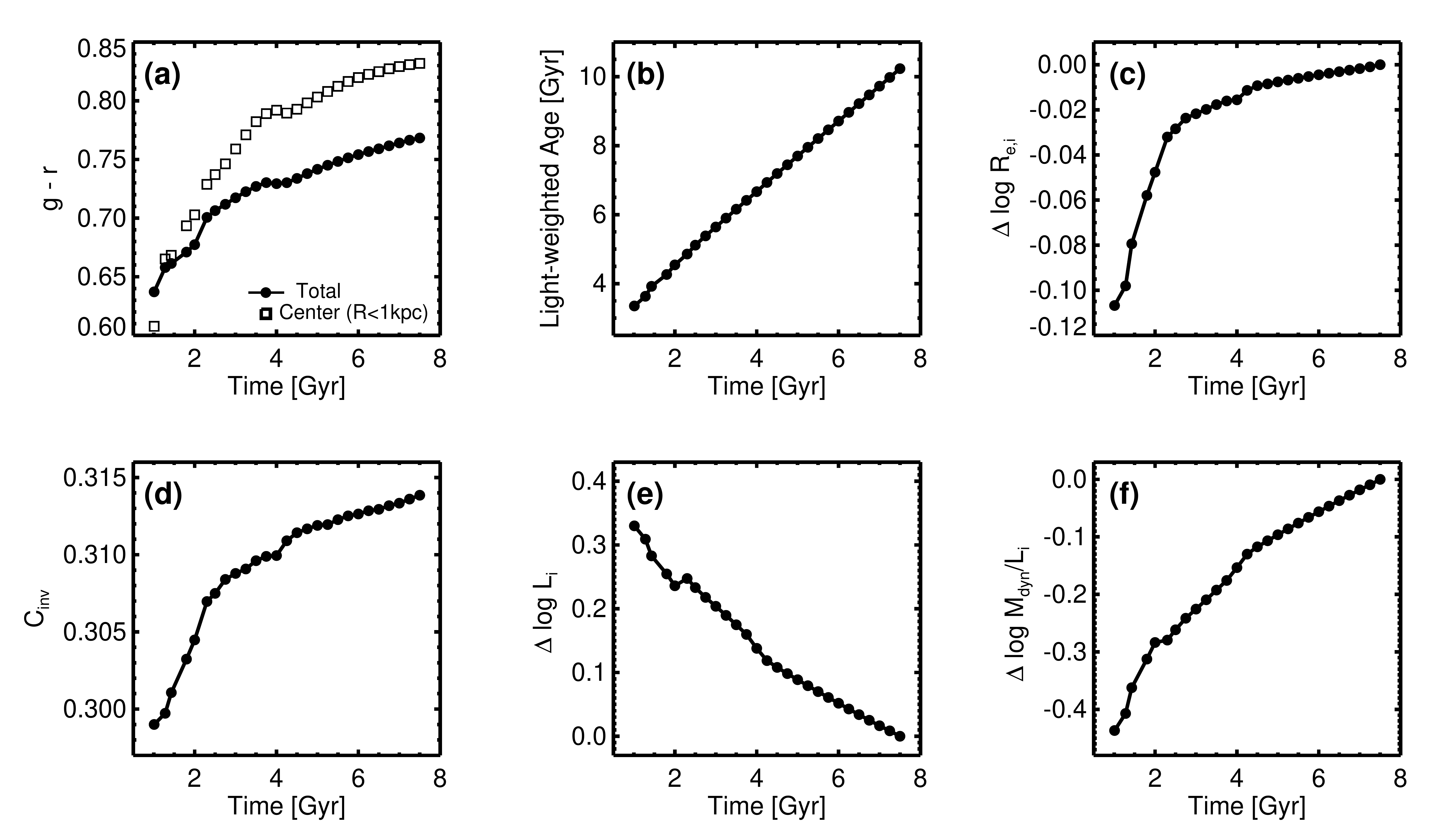}  
\centering
\caption{Properties of a galaxy that experienced a gas-rich major merger as a function of $t$, where $t$ is time since the instantaneous starburst by the merger. (a) Total and central $g-r$ color within 1 kpc. (b) Light-weighted age ($i$-band luminosity-weighted age). (c) Half-light radius in $i$ band ($\log R_{e.i}$). (d) Inverse concentration index in $i$ band ($C_\mathrm{inv}$). (e) Luminosity in $i$ band  ($\log L_i$). (f) $\log M_\mathrm{dyn}/L_i$. For $\log R_{e.i}$, $\log L_i$, and $\log M_\mathrm{dyn}/L_i$, relative differences are shown ($t=7.5$ Gyr is the reference point). 
\label{fig:sim}}
\end{figure*}

We generated a simple model of a typical galaxy that experienced a gas-rich major merger. Specifically, we created a post-merger model that has two-component stellar populations (one is an old stellar population that already existed in progenitors, and the other is a newly formed one in the merger process), whose properties such as color, age, metallicity, and luminosity were produced using \citet{Bruzual2003} stellar population models with a \citet{Chabrier2003} IMF.  Additionally, we assigned surface brightness profiles, which follow the S\'{e}rsic profile, to the old and young components of the model, so that the model has structure properties such as compactness and the half-light radius. As described in the following sentences, all the specific stellar population properties of the components and their surface brightness profiles are based on simulation studies for remnants of gas-rich major mergers \citep{Hopkins2008b,Hopkins2009}. After generating the post-merger model, we compared its properties such as color, compactness, and $M_\mathrm{dyn}/L$ to those of compact ETGs in our study. By doing so, we verify the connection between gas-rich major mergers and compact ETGs.

Gas consumption by starburst and feedbacks from supermassive black holes during the merger process rapidly quench star formation activities within $\sim1$ Gyr \citep{Hopkins2008c}, so that galaxies become red without extended star formation \citep{Peng2010,Brennan2015}. Thus, in the simple model, we assume that all the stars form instantaneously by the merger and the final mass fraction of the newly formed stars is $10\%$, which is a typical value for gas-rich merger remnants \citep{Hopkins2008b,Hopkins2009}. We set surface brightness profiles of the newly formed stars to follow the exponential profile whose half-light radius is 0.7 kpc, which is a typical property of compact extra light components of young stellar populations in ETGs produced by gas-rich major mergers \citep{Hopkins2008b,Hopkins2009}.

Old stars that already existed in progenitors of the merger remnant are set to follow the S\'{e}rsic profile of which S\'{e}rsic index and half-light radius are 2 and 5.1 kpc, respectively, in the post-merger galaxy. This is also a typical profile of the outer components of old stellar populations in ETGs \citep{Hopkins2008b,Hopkins2009}. We assumed that ages of old stars are 3 Gyr at the moment of the instantaneous starburst by the merger, following age gradients for simulated elliptical galaxies that experienced gas-rich mergers shown in \citet{Hopkins2009}. 

As a reflection of the typical metallicity gradient of ETGs \citep{Hopkins2009,Kim2013}, we set the young stellar population formed at the inner part of the merger remnant to have high metallicity of $Z=0.05$, while the old stellar population in the outer part is set to have low metallicity of $Z=0.008$.

We examine the properties of the gas-rich merger remnant as a function of time since 1 Gyr after the instantaneous starburst by the merger. We note that merger remnants are in general fully relaxed $\sim1$--$2$ Gyr after the final merger of galaxies \citep{Hopkins2008b,Hopkins2009}. We only permit passive evolution of the stellar populations in the simple merger remnant model, while all the other properties of the model galaxy are assumed to be constant in the passive evolution. 

The results are presented in Figure \ref{fig:sim}, which shows several properties of the merger remnant as a function of time since the initial starburst by the merger (hereafter $t$) in a range of $1$ Gyr $\le t\le7.5$ Gyr. Panels (a) and (b) of Figure \ref{fig:sim} show that the post-merger galaxy at $t\lesssim3$ Gyr has blue total color of $g-r\sim0.68$ and light-weighted age of $\lesssim5.5$ Gyr. On the other hand, those at $t\gtrsim5$ Gyr have redder total color of $g-r\gtrsim0.75$ and light-weighted age of $\gtrsim8$ Gyr. The central color within 1 kpc at $t\lesssim3$ Gyr is comparable to or only slightly redder than the total color. However, at $t\gtrsim5$ Gyr, the central color escalates to $g-r\sim0.84$ that is far redder than the total color, due to the effect of high metallicity in the central region of the merger remnant. 

As time goes by, the brightness of the inner young component of the merger remnant more rapidly decreases than the already old outer component. Thus, as shown in panel (c) of Figure \ref{fig:sim}, a half-light radius of the post-merger galaxy naturally increases as a function of $t$, particularly rapidly at $t\lesssim3$ Gyr, so that the post-merger galaxy at $t\lesssim3$ Gyr has a $\sim0.1$ dex ($\sim25\%$) smaller half-light radius than that at $t\gtrsim5$ Gyr. Accordingly, $C_\mathrm{inv}$ in $i$ band also increases as a function of $t$ (panel (d) of Figure \ref{fig:sim}). However, the increment of $C_\mathrm{inv}$ is very minor, which is $\sim0.015$ in 6 Gyr, so that $C_\mathrm{inv}$ is always less than 0.33 in the entire $t$ range we probe. Thus, the merger remnant formed by the gas-rich major merger always satisfies the criterion for compact galaxies used in this study ($C_\mathrm{inv}<0.33$) regardless of $t$. We note that the outer old component alone has $C_\mathrm{inv}=0.344$, but the newly formed inner component by the merger makes the merger remnant more compact. 

$L_i$ of the merger remnant naturally decreases as a function of $t$ ($\sim0.30$ dex in 7 Gyr) owing to its passive evolution as shown in panel (e) of Figure \ref{fig:sim}. We calculated $M_\mathrm{dyn}$ of the post-merger galaxy using the same method applied to observational data (Equation \ref{eq:dym}). Since the mass distribution in the post-merger galaxy after the completion or relaxation of the merger is assumed to be invariable, the computed $M_\mathrm{dyn}$ at a certain $t$ is only affected by the change in the half-light radius of the merger remnant. Combining evolutions of $M_\mathrm{dyn}$ and $L_i$, we examine $M_\mathrm{dyn}/L_i$ of the post-merger galaxy as a funtion of $t$, which is shown in panel (f) of Figure \ref{fig:sim}. The figure shows that the post-merger galaxy at $t\lesssim3$ Gyr has $\sim0.2$--$0.3$ dex smaller $M_\mathrm{dyn}/L_i$ than that at $t\gtrsim5$ Gyr, due to its compact size and high luminosity.

The simple model of the galaxy that experienced a gas-rich major merger well confirms that a gas-rich major merger can make blue color particularly in the galaxy central region, compact light distribution, and low $M_\mathrm{dyn}/L$ for a few gigayears after the merger. In terms of galaxy properties such as total color, central color, age, and compactness, the model galaxies at $t\lesssim5$ Gyr are similar to compact young ETGs ($C_\mathrm{inv}<0.33$ and $\Delta (g-r)_\mathrm{P}<-0.03$) with low $\log M_\mathrm{dyn}/L\,(\lesssim0.4)$, while those at  $t\gtrsim5$ Gyr are similar to compact old ETGs ($C_\mathrm{inv}<0.33$ and $\Delta (g-r)_\mathrm{P}\ge0.00$) in our study (see Figures \ref{fig:mlcon} and \ref{fig:cml}). This implies that compact young ETGs with low $M_\mathrm{dyn}/L$ and blue central color are galaxies that have experienced recent gas-rich major mergers, and they can naturally become compact old ETGs with the passage of time.
\\

\section{Summary}\label{sec:summary}

Using $16,283$ ETGs with $M_r\le-19.5$ and $0.025\le z<0.055$ from SDSS data, we examine age and internal structure dependence of the scatter in the FP. We use $C_\mathrm{inv}$ for the structure parameter of ETGs, which denotes how compact galaxy light distribution is. For an age indicator for ETGs, we use $\Delta (g-r)_\mathrm{P}$ that is defined in the three-dimensional parameter space of $g-r$, $M_r$, and $\log\sigma_0$. The method of $\chi^2$ minimization after trimming outliers \citep{Cappellari2013} is used to fit FP to data. The main conclusions of this study are summarized as follows.
\begin{enumerate}
\item We confirm that the scatter of the FP is smaller in the redder band (the smallest in $z$ band).
\item We find that the size of the scatter in the FP depends on galaxy age: old ETGs with age $\gtrsim9$ Gyrs have a smaller intrinsic scatter ($\sim0.06$ dex; $\sim14\%$) in the FP than young ETGs with age $\lesssim6$ Gyrs ($\sim0.075$ dex; $\sim17\%$).
\item For young ETGs, we find that less compact ETGs have a smaller scatter in the FP ($\sim0.065$ dex; $\sim15\%$) than more compact ones ($\sim0.10$ dex; $\sim23\%$). On the other hand, in the case of old ETGs, the scatter in the FP does not depend on the compactness of galaxy structure. 
\item We find that more compact and younger ETGs have a larger scatter in the FP than other ETG subpopulations. This large scatter in compact young ETGs is caused by ETGs that have excessive light for a given mass (low $M_\mathrm{dyn}/L$) and blue color in the centers.
\item A simple model of a galaxy that experienced a gas-rich major merger shows that gas-rich major mergers can make more compact structure, lower $M_\mathrm{dyn}/L$, and younger stellar populations especially in the central regions of ETGs. Thus, we conclude that compact young ETGs with low $M_\mathrm{dyn}/L$ and central blue color are the galaxies that have experienced recent gas-rich major mergers.
\end{enumerate}

We suggest that compact young ETGs with low $M_\mathrm{dyn}/L$ and central blue color have experienced recent gas-rich major mergers. Further studies on finding more direct evidence for mergers such as tidal features using deep images can strengthen our argument. In addition, our results on FPs that have reduced scatters are expected to be used for more accurate distance measurements in future studies.

\acknowledgments
We thank the anonymous referee for helpful comments that improved the content of the paper. This work was supported by a KIAS Individual Grant PG076301 at the Korea Institute for Advanced Study.

\appendix

\section{Intrinsic Scatters in the FP and Their Coefficients}\label{AppendixTable}
All the values of intrinsic scatters in the FPs and their coefficients in Section \ref{sec:results:scatter} are summarized in Tables \ref{tb:g}, \ref{tb:r}, \ref{tb:i}, and \ref{tb:z} for $g$, $r$, $i$, and $z$ bands, respectively.

\begin{deluxetable*}{rcccccc}
\tablecaption{Intrinsic Scatters in the FPs and Their Coefficients ($g$ Band)\label{tb:g}}
\tabletypesize{\scriptsize}
\tablehead{\colhead{Category} & \colhead{$a$} & \colhead{$b$} & \colhead{$c$} & \colhead{$\varepsilon$} & \colhead{rms} & \colhead{$N$}
}
\startdata
All&$0.966\pm0.006$&$0.287\pm0.001$&$-7.578\pm0.025$&$0.087\pm0.001$&$0.090$&$16283$\\
\hline
$\Delta (g-r)_\mathrm{P}<-0.03$&$1.165\pm0.014$&$0.229\pm0.002$&$-6.747\pm0.044$&$0.081\pm0.001$&$0.085$&$ 3972$\\
$0.00\leq \Delta (g-r)_\mathrm{P} < 0.075$&$1.117\pm0.008$&$0.302\pm0.001$&$-8.240\pm0.032$&$0.071\pm0.001$&$0.074$&$ 7600$\\
\hline
$\Delta (g-r)_\mathrm{P}<0.02$, $C_\mathrm{inv}<0.33$&$1.055\pm0.012$&$0.243\pm0.002$&$-6.848\pm0.046$&$0.089\pm0.001$&$0.092$&$ 5756$\\
$\Delta (g-r)_\mathrm{P}<0.02$, $C_\mathrm{inv}\geq 0.33$&$1.188\pm0.012$&$0.270\pm0.001$&$-7.678\pm0.036$&$0.075\pm0.001$&$0.080$&$ 5410$\\
$\Delta (g-r)_\mathrm{P}\geq 0.02$, $C_\mathrm{inv}<0.33$&$1.110\pm0.016$&$0.310\pm0.002$&$-8.405\pm0.060$&$0.071\pm0.001$&$0.074$&$ 2750$\\
$\Delta (g-r)_\mathrm{P}\geq 0.02$, $C_\mathrm{inv}\geq 0.33$&$1.068\pm0.015$&$0.306\pm0.002$&$-8.245\pm0.054$&$0.068\pm0.001$&$0.072$&$ 2367$\\
\hline
$-0.100\leq \Delta (g-r)_\mathrm{P}<-0.075$&$1.236\pm0.036$&$0.217\pm0.003$&$-6.609\pm0.100$&$0.076\pm0.003$&$0.081$&$  584$\\
$-0.075\leq \Delta (g-r)_\mathrm{P}<-0.050$&$1.166\pm0.024$&$0.225\pm0.003$&$-6.651\pm0.075$&$0.075\pm0.002$&$0.079$&$ 1235$\\
$-0.050\leq \Delta (g-r)_\mathrm{P}<-0.025$&$1.239\pm0.016$&$0.239\pm0.002$&$-7.138\pm0.051$&$0.070\pm0.001$&$0.075$&$ 2404$\\
$-0.025\leq \Delta (g-r)_\mathrm{P}<0.000$&$1.225\pm0.012$&$0.264\pm0.002$&$-7.671\pm0.044$&$0.067\pm0.001$&$0.072$&$ 3567$\\
$0.000\leq \Delta (g-r)_\mathrm{P}<0.025$&$1.201\pm0.011$&$0.285\pm0.002$&$-8.078\pm0.043$&$0.066\pm0.001$&$0.070$&$ 3745$\\
$0.025\leq \Delta (g-r)_\mathrm{P}<0.050$&$1.152\pm0.014$&$0.298\pm0.002$&$-8.259\pm0.051$&$0.067\pm0.001$&$0.070$&$ 2602$\\
$0.050\leq \Delta (g-r)_\mathrm{P}<0.075$&$1.149\pm0.020$&$0.317\pm0.003$&$-8.648\pm0.074$&$0.066\pm0.001$&$0.070$&$ 1253$\\
$0.075\leq \Delta (g-r)_\mathrm{P}<0.100$&$1.105\pm0.037$&$0.316\pm0.005$&$-8.540\pm0.140$&$0.075\pm0.003$&$0.077$&$  451$\\
\hline
$\Delta (g-r)_\mathrm{P}<0.02, 0.27\leq C_\mathrm{inv}<0.29$&$0.923\pm0.046$&$0.182\pm0.009$&$-5.340\pm0.203$&$0.100\pm0.004$&$0.102$&$  490$\\
$\Delta (g-r)_\mathrm{P}<0.02, 0.29\leq C_\mathrm{inv}<0.31$&$1.052\pm0.019$&$0.232\pm0.004$&$-6.616\pm0.080$&$0.089\pm0.001$&$0.091$&$ 2356$\\
$\Delta (g-r)_\mathrm{P}<0.02, 0.31\leq C_\mathrm{inv}<0.33$&$1.107\pm0.017$&$0.252\pm0.002$&$-7.153\pm0.057$&$0.084\pm0.001$&$0.087$&$ 2862$\\
$\Delta (g-r)_\mathrm{P}<0.02, 0.33\leq C_\mathrm{inv}<0.35$&$1.181\pm0.020$&$0.265\pm0.002$&$-7.566\pm0.060$&$0.080\pm0.001$&$0.084$&$ 2124$\\
$\Delta (g-r)_\mathrm{P}<0.02, 0.35\leq C_\mathrm{inv}<0.37$&$1.207\pm0.022$&$0.268\pm0.002$&$-7.686\pm0.063$&$0.072\pm0.001$&$0.077$&$ 1599$\\
$\Delta (g-r)_\mathrm{P}<0.02, 0.37\leq C_\mathrm{inv}<0.39$&$1.244\pm0.031$&$0.264\pm0.003$&$-7.671\pm0.086$&$0.070\pm0.002$&$0.077$&$  864$\\
$\Delta (g-r)_\mathrm{P}<0.02, 0.39\leq C_\mathrm{inv}<0.41$&$1.336\pm0.046$&$0.277\pm0.005$&$-8.132\pm0.121$&$0.069\pm0.003$&$0.077$&$  466$\\
$\Delta (g-r)_\mathrm{P}\geq 0.02, 0.27\leq C_\mathrm{inv}<0.29$&$1.136\pm0.049$&$0.289\pm0.010$&$-8.064\pm0.229$&$0.067\pm0.004$&$0.069$&$  257$\\
$\Delta (g-r)_\mathrm{P}\geq 0.02, 0.29\leq C_\mathrm{inv}<0.31$&$1.098\pm0.025$&$0.309\pm0.004$&$-8.361\pm0.104$&$0.072\pm0.002$&$0.074$&$ 1153$\\
$\Delta (g-r)_\mathrm{P}\geq 0.02, 0.31\leq C_\mathrm{inv}<0.33$&$1.132\pm0.024$&$0.313\pm0.003$&$-8.522\pm0.080$&$0.072\pm0.001$&$0.075$&$ 1308$\\
$\Delta (g-r)_\mathrm{P}\geq 0.02, 0.33\leq C_\mathrm{inv}<0.35$&$1.053\pm0.025$&$0.309\pm0.003$&$-8.254\pm0.084$&$0.068\pm0.002$&$0.071$&$ 1028$\\
$\Delta (g-r)_\mathrm{P}\geq 0.02, 0.35\leq C_\mathrm{inv}<0.37$&$0.970\pm0.028$&$0.306\pm0.003$&$-8.016\pm0.096$&$0.068\pm0.002$&$0.071$&$  747$\\
$\Delta (g-r)_\mathrm{P}\geq 0.02, 0.37\leq C_\mathrm{inv}<0.39$&$1.139\pm0.038$&$0.310\pm0.004$&$-8.469\pm0.129$&$0.064\pm0.003$&$0.069$&$  337$\\
$\Delta (g-r)_\mathrm{P}\geq 0.02, 0.39\leq C_\mathrm{inv}<0.41$&$1.210\pm0.065$&$0.302\pm0.006$&$-8.463\pm0.208$&$0.070\pm0.005$&$0.072$&$  163$\\
\enddata
\tablecomments{$a$, $b$, and $c$ are coefficients of the FP (Equation \ref{eq:fp}), while $\varepsilon$ is the intrinsic scatter in the FP (see Equation \ref{eq:chi}). The rms is observed root-mean-square values. $N$ is the number of ETGs in each category.
}
\end{deluxetable*}

\begin{deluxetable*}{rcccccc}
\tablecaption{Intrinsic Scatters in the FPs and Their Coefficients ($r$ Band)\label{tb:r}}
\tabletypesize{\scriptsize}
\tablehead{\colhead{Category} & \colhead{$a$} & \colhead{$b$} & \colhead{$c$} & \colhead{$\varepsilon$} & \colhead{rms} & \colhead{$N$}
}
\startdata
All&$1.014\pm0.006$&$0.293\pm0.001$&$-7.564\pm0.024$&$0.085\pm0.001$&$0.087$&$16283$\\
\hline
$\Delta (g-r)_\mathrm{P}<-0.03$&$1.184\pm0.014$&$0.236\pm0.002$&$-6.754\pm0.045$&$0.081\pm0.001$&$0.085$&$ 3972$\\
$0.00\leq \Delta (g-r)_\mathrm{P} < 0.075$&$1.138\pm0.008$&$0.307\pm0.001$&$-8.151\pm0.031$&$0.070\pm0.001$&$0.073$&$ 7600$\\
\hline
$\Delta (g-r)_\mathrm{P}<0.02$, $C_\mathrm{inv}<0.33$&$1.083\pm0.012$&$0.253\pm0.002$&$-6.925\pm0.045$&$0.087\pm0.001$&$0.089$&$ 5756$\\
$\Delta (g-r)_\mathrm{P}<0.02$, $C_\mathrm{inv}\geq 0.33$&$1.235\pm0.012$&$0.274\pm0.001$&$-7.650\pm0.036$&$0.073\pm0.001$&$0.079$&$ 5410$\\
$\Delta (g-r)_\mathrm{P}\geq 0.02$, $C_\mathrm{inv}<0.33$&$1.132\pm0.015$&$0.317\pm0.002$&$-8.339\pm0.057$&$0.071\pm0.001$&$0.073$&$ 2750$\\
$\Delta (g-r)_\mathrm{P}\geq 0.02$, $C_\mathrm{inv}\geq 0.33$&$1.124\pm0.015$&$0.313\pm0.002$&$-8.248\pm0.054$&$0.068\pm0.001$&$0.072$&$ 2367$\\
\hline
$-0.100\leq \Delta (g-r)_\mathrm{P}<-0.075$&$1.249\pm0.036$&$0.221\pm0.004$&$-6.568\pm0.108$&$0.078\pm0.003$&$0.083$&$  584$\\
$-0.075\leq \Delta (g-r)_\mathrm{P}<-0.050$&$1.177\pm0.024$&$0.231\pm0.003$&$-6.634\pm0.079$&$0.076\pm0.002$&$0.080$&$ 1235$\\
$-0.050\leq \Delta (g-r)_\mathrm{P}<-0.025$&$1.253\pm0.016$&$0.245\pm0.002$&$-7.114\pm0.053$&$0.071\pm0.001$&$0.076$&$ 2404$\\
$-0.025\leq \Delta (g-r)_\mathrm{P}<0.000$&$1.238\pm0.012$&$0.271\pm0.002$&$-7.630\pm0.044$&$0.068\pm0.001$&$0.073$&$ 3567$\\
$0.000\leq \Delta (g-r)_\mathrm{P}<0.025$&$1.214\pm0.011$&$0.290\pm0.002$&$-7.975\pm0.043$&$0.067\pm0.001$&$0.071$&$ 3745$\\
$0.025\leq \Delta (g-r)_\mathrm{P}<0.050$&$1.161\pm0.014$&$0.306\pm0.002$&$-8.182\pm0.051$&$0.068\pm0.001$&$0.071$&$ 2602$\\
$0.050\leq \Delta (g-r)_\mathrm{P}<0.075$&$1.162\pm0.020$&$0.325\pm0.003$&$-8.570\pm0.072$&$0.065\pm0.001$&$0.068$&$ 1253$\\
$0.075\leq \Delta (g-r)_\mathrm{P}<0.100$&$1.112\pm0.036$&$0.325\pm0.005$&$-8.469\pm0.134$&$0.073\pm0.003$&$0.075$&$  451$\\
\hline
$\Delta (g-r)_\mathrm{P}<0.02, 0.27\leq C_\mathrm{inv}<0.29$&$0.928\pm0.046$&$0.204\pm0.010$&$-5.663\pm0.219$&$0.102\pm0.004$&$0.103$&$  490$\\
$\Delta (g-r)_\mathrm{P}<0.02, 0.29\leq C_\mathrm{inv}<0.31$&$1.077\pm0.019$&$0.245\pm0.004$&$-6.752\pm0.080$&$0.087\pm0.001$&$0.089$&$ 2356$\\
$\Delta (g-r)_\mathrm{P}<0.02, 0.31\leq C_\mathrm{inv}<0.33$&$1.139\pm0.016$&$0.259\pm0.002$&$-7.160\pm0.056$&$0.083\pm0.001$&$0.085$&$ 2862$\\
$\Delta (g-r)_\mathrm{P}<0.02, 0.33\leq C_\mathrm{inv}<0.35$&$1.223\pm0.019$&$0.269\pm0.002$&$-7.541\pm0.060$&$0.078\pm0.001$&$0.082$&$ 2124$\\
$\Delta (g-r)_\mathrm{P}<0.02, 0.35\leq C_\mathrm{inv}<0.37$&$1.256\pm0.022$&$0.271\pm0.002$&$-7.647\pm0.062$&$0.070\pm0.001$&$0.076$&$ 1599$\\
$\Delta (g-r)_\mathrm{P}<0.02, 0.37\leq C_\mathrm{inv}<0.39$&$1.291\pm0.030$&$0.265\pm0.003$&$-7.584\pm0.086$&$0.070\pm0.002$&$0.076$&$  864$\\
$\Delta (g-r)_\mathrm{P}<0.02, 0.39\leq C_\mathrm{inv}<0.41$&$1.388\pm0.044$&$0.284\pm0.005$&$-8.157\pm0.122$&$0.068\pm0.004$&$0.075$&$  466$\\
$\Delta (g-r)_\mathrm{P}\geq 0.02, 0.27\leq C_\mathrm{inv}<0.29$&$1.165\pm0.048$&$0.301\pm0.010$&$-8.127\pm0.227$&$0.066\pm0.003$&$0.068$&$  257$\\
$\Delta (g-r)_\mathrm{P}\geq 0.02, 0.29\leq C_\mathrm{inv}<0.31$&$1.121\pm0.025$&$0.318\pm0.004$&$-8.341\pm0.099$&$0.070\pm0.002$&$0.072$&$ 1153$\\
$\Delta (g-r)_\mathrm{P}\geq 0.02, 0.31\leq C_\mathrm{inv}<0.33$&$1.171\pm0.023$&$0.320\pm0.003$&$-8.476\pm0.077$&$0.070\pm0.002$&$0.073$&$ 1308$\\
$\Delta (g-r)_\mathrm{P}\geq 0.02, 0.33\leq C_\mathrm{inv}<0.35$&$1.109\pm0.025$&$0.313\pm0.003$&$-8.218\pm0.085$&$0.069\pm0.002$&$0.072$&$ 1028$\\
$\Delta (g-r)_\mathrm{P}\geq 0.02, 0.35\leq C_\mathrm{inv}<0.37$&$1.033\pm0.028$&$0.313\pm0.003$&$-8.046\pm0.097$&$0.068\pm0.002$&$0.071$&$  747$\\
$\Delta (g-r)_\mathrm{P}\geq 0.02, 0.37\leq C_\mathrm{inv}<0.39$&$1.207\pm0.040$&$0.315\pm0.005$&$-8.463\pm0.135$&$0.066\pm0.003$&$0.071$&$  337$\\
$\Delta (g-r)_\mathrm{P}\geq 0.02, 0.39\leq C_\mathrm{inv}<0.41$&$1.255\pm0.065$&$0.311\pm0.007$&$-8.472\pm0.212$&$0.070\pm0.005$&$0.075$&$  163$\\
\enddata
\tablecomments{$a$, $b$, and $c$ are coefficients of the FP (Equation \ref{eq:fp}), while $\varepsilon$ is the intrinsic scatter in the FP (see Equation \ref{eq:chi}). The rms is observed root-mean-square values. $N$ is the number of ETGs in each category.
}
\end{deluxetable*}

\begin{deluxetable*}{rcccccc}
\tablecaption{Intrinsic Scatters in the FPs and Their Coefficients ($i$ Band)\label{tb:i}}
\tabletypesize{\scriptsize}
\tablehead{\colhead{Category} & \colhead{$a$} & \colhead{$b$} & \colhead{$c$} & \colhead{$\varepsilon$} & \colhead{rms} & \colhead{$N$}
}
\startdata
All&$1.067\pm0.006$&$0.299\pm0.001$&$-7.691\pm0.024$&$0.082\pm0.001$&$0.085$&$16283$\\
\hline
$\Delta (g-r)_\mathrm{P}<-0.03$&$1.212\pm0.014$&$0.242\pm0.002$&$-6.864\pm0.046$&$0.081\pm0.001$&$0.085$&$ 3972$\\
$0.00\leq \Delta (g-r)_\mathrm{P} < 0.075$&$1.176\pm0.008$&$0.314\pm0.001$&$-8.255\pm0.031$&$0.069\pm0.001$&$0.072$&$ 7600$\\
\hline
$\Delta (g-r)_\mathrm{P}<0.02$, $C_\mathrm{inv}<0.33$&$1.120\pm0.011$&$0.262\pm0.002$&$-7.085\pm0.045$&$0.086\pm0.001$&$0.089$&$ 5756$\\
$\Delta (g-r)_\mathrm{P}<0.02$, $C_\mathrm{inv}\geq 0.33$&$1.273\pm0.011$&$0.279\pm0.001$&$-7.744\pm0.036$&$0.072\pm0.001$&$0.078$&$ 5410$\\
$\Delta (g-r)_\mathrm{P}\geq 0.02$, $C_\mathrm{inv}<0.33$&$1.165\pm0.015$&$0.325\pm0.002$&$-8.453\pm0.056$&$0.069\pm0.001$&$0.071$&$ 2750$\\
$\Delta (g-r)_\mathrm{P}\geq 0.02$, $C_\mathrm{inv}\geq 0.33$&$1.159\pm0.015$&$0.320\pm0.002$&$-8.327\pm0.054$&$0.067\pm0.001$&$0.071$&$ 2367$\\
\hline
$-0.100\leq \Delta (g-r)_\mathrm{P}<-0.075$&$1.277\pm0.036$&$0.228\pm0.004$&$-6.690\pm0.110$&$0.078\pm0.003$&$0.083$&$  584$\\
$-0.075\leq \Delta (g-r)_\mathrm{P}<-0.050$&$1.202\pm0.025$&$0.236\pm0.003$&$-6.719\pm0.082$&$0.077\pm0.002$&$0.081$&$ 1235$\\
$-0.050\leq \Delta (g-r)_\mathrm{P}<-0.025$&$1.281\pm0.016$&$0.251\pm0.002$&$-7.207\pm0.054$&$0.072\pm0.001$&$0.076$&$ 2404$\\
$-0.025\leq \Delta (g-r)_\mathrm{P}<0.000$&$1.266\pm0.012$&$0.278\pm0.002$&$-7.721\pm0.045$&$0.068\pm0.001$&$0.073$&$ 3567$\\
$0.000\leq \Delta (g-r)_\mathrm{P}<0.025$&$1.244\pm0.011$&$0.297\pm0.002$&$-8.067\pm0.043$&$0.067\pm0.001$&$0.071$&$ 3745$\\
$0.025\leq \Delta (g-r)_\mathrm{P}<0.050$&$1.193\pm0.014$&$0.313\pm0.002$&$-8.276\pm0.051$&$0.067\pm0.001$&$0.070$&$ 2602$\\
$0.050\leq \Delta (g-r)_\mathrm{P}<0.075$&$1.182\pm0.020$&$0.333\pm0.003$&$-8.637\pm0.072$&$0.064\pm0.001$&$0.067$&$ 1253$\\
$0.075\leq \Delta (g-r)_\mathrm{P}<0.100$&$1.123\pm0.035$&$0.332\pm0.005$&$-8.495\pm0.132$&$0.071\pm0.003$&$0.074$&$  451$\\
\hline
$\Delta (g-r)_\mathrm{P}<0.02, 0.27\leq C_\mathrm{inv}<0.29$&$0.954\pm0.046$&$0.224\pm0.011$&$-6.021\pm0.226$&$0.102\pm0.004$&$0.104$&$  490$\\
$\Delta (g-r)_\mathrm{P}<0.02, 0.29\leq C_\mathrm{inv}<0.31$&$1.113\pm0.018$&$0.255\pm0.004$&$-6.940\pm0.080$&$0.086\pm0.001$&$0.088$&$ 2356$\\
$\Delta (g-r)_\mathrm{P}<0.02, 0.31\leq C_\mathrm{inv}<0.33$&$1.177\pm0.016$&$0.266\pm0.002$&$-7.279\pm0.056$&$0.081\pm0.001$&$0.084$&$ 2862$\\
$\Delta (g-r)_\mathrm{P}<0.02, 0.33\leq C_\mathrm{inv}<0.35$&$1.261\pm0.019$&$0.273\pm0.002$&$-7.610\pm0.060$&$0.077\pm0.001$&$0.082$&$ 2124$\\
$\Delta (g-r)_\mathrm{P}<0.02, 0.35\leq C_\mathrm{inv}<0.37$&$1.287\pm0.021$&$0.278\pm0.003$&$-7.751\pm0.062$&$0.068\pm0.001$&$0.074$&$ 1599$\\
$\Delta (g-r)_\mathrm{P}<0.02, 0.37\leq C_\mathrm{inv}<0.39$&$1.332\pm0.030$&$0.271\pm0.003$&$-7.705\pm0.087$&$0.068\pm0.002$&$0.075$&$  864$\\
$\Delta (g-r)_\mathrm{P}<0.02, 0.39\leq C_\mathrm{inv}<0.41$&$1.419\pm0.043$&$0.291\pm0.005$&$-8.278\pm0.124$&$0.067\pm0.003$&$0.075$&$  466$\\
$\Delta (g-r)_\mathrm{P}\geq 0.02, 0.27\leq C_\mathrm{inv}<0.29$&$1.192\pm0.049$&$0.314\pm0.010$&$-8.325\pm0.226$&$0.067\pm0.004$&$0.069$&$  257$\\
$\Delta (g-r)_\mathrm{P}\geq 0.02, 0.29\leq C_\mathrm{inv}<0.31$&$1.153\pm0.024$&$0.327\pm0.004$&$-8.461\pm0.097$&$0.068\pm0.002$&$0.070$&$ 1153$\\
$\Delta (g-r)_\mathrm{P}\geq 0.02, 0.31\leq C_\mathrm{inv}<0.33$&$1.208\pm0.022$&$0.326\pm0.003$&$-8.554\pm0.075$&$0.068\pm0.002$&$0.071$&$ 1308$\\
$\Delta (g-r)_\mathrm{P}\geq 0.02, 0.33\leq C_\mathrm{inv}<0.35$&$1.153\pm0.025$&$0.319\pm0.003$&$-8.301\pm0.085$&$0.068\pm0.002$&$0.071$&$ 1028$\\
$\Delta (g-r)_\mathrm{P}\geq 0.02, 0.35\leq C_\mathrm{inv}<0.37$&$1.077\pm0.027$&$0.322\pm0.003$&$-8.188\pm0.096$&$0.065\pm0.002$&$0.069$&$  747$\\
$\Delta (g-r)_\mathrm{P}\geq 0.02, 0.37\leq C_\mathrm{inv}<0.39$&$1.233\pm0.039$&$0.322\pm0.005$&$-8.532\pm0.133$&$0.064\pm0.003$&$0.070$&$  337$\\
$\Delta (g-r)_\mathrm{P}\geq 0.02, 0.39\leq C_\mathrm{inv}<0.41$&$1.277\pm0.064$&$0.319\pm0.007$&$-8.563\pm0.208$&$0.068\pm0.005$&$0.073$&$  163$\\
\enddata
\tablecomments{$a$, $b$, and $c$ are coefficients of the FP (Equation \ref{eq:fp}), while $\varepsilon$ is the intrinsic scatter in the FP (see Equation \ref{eq:chi}). The rms is observed root-mean-square values. $N$ is the number of ETGs in each category.
}
\end{deluxetable*}

\begin{deluxetable*}{rcccccc}
\tablecaption{Intrinsic Scatters in the FPs and Their Coefficients ($z$ Band)\label{tb:z}}
\tabletypesize{\scriptsize}
\tablehead{\colhead{Category} & \colhead{$a$} & \colhead{$b$} & \colhead{$c$} & \colhead{$\varepsilon$} & \colhead{rms} & \colhead{$N$}
}
\startdata
All&$1.086\pm0.006$&$0.306\pm0.001$&$-7.769\pm0.024$&$0.078\pm0.001$&$0.085$&$16283$\\
\hline
$\Delta (g-r)_\mathrm{P}<-0.03$&$1.195\pm0.014$&$0.250\pm0.002$&$-6.913\pm0.047$&$0.079\pm0.001$&$0.086$&$ 3972$\\
$0.00\leq \Delta (g-r)_\mathrm{P} < 0.075$&$1.199\pm0.008$&$0.324\pm0.001$&$-8.383\pm0.031$&$0.063\pm0.001$&$0.073$&$ 7600$\\
\hline
$\Delta (g-r)_\mathrm{P}<0.02$, $C_\mathrm{inv}<0.33$&$1.112\pm0.011$&$0.272\pm0.002$&$-7.180\pm0.045$&$0.083\pm0.001$&$0.089$&$ 5756$\\
$\Delta (g-r)_\mathrm{P}<0.02$, $C_\mathrm{inv}\geq 0.33$&$1.277\pm0.011$&$0.288\pm0.001$&$-7.821\pm0.036$&$0.067\pm0.001$&$0.078$&$ 5410$\\
$\Delta (g-r)_\mathrm{P}\geq 0.02$, $C_\mathrm{inv}<0.33$&$1.186\pm0.015$&$0.335\pm0.002$&$-8.566\pm0.056$&$0.062\pm0.001$&$0.068$&$ 2750$\\
$\Delta (g-r)_\mathrm{P}\geq 0.02$, $C_\mathrm{inv}\geq 0.33$&$1.186\pm0.015$&$0.330\pm0.002$&$-8.472\pm0.055$&$0.062\pm0.001$&$0.078$&$ 2367$\\
\hline
$-0.100\leq \Delta (g-r)_\mathrm{P}<-0.075$&$1.242\pm0.038$&$0.237\pm0.004$&$-6.722\pm0.117$&$0.080\pm0.003$&$0.088$&$  584$\\
$-0.075\leq \Delta (g-r)_\mathrm{P}<-0.050$&$1.186\pm0.025$&$0.245\pm0.003$&$-6.794\pm0.084$&$0.076\pm0.002$&$0.082$&$ 1235$\\
$-0.050\leq \Delta (g-r)_\mathrm{P}<-0.025$&$1.255\pm0.016$&$0.261\pm0.002$&$-7.273\pm0.057$&$0.070\pm0.001$&$0.078$&$ 2404$\\
$-0.025\leq \Delta (g-r)_\mathrm{P}<0.000$&$1.259\pm0.012$&$0.290\pm0.002$&$-7.839\pm0.046$&$0.065\pm0.001$&$0.073$&$ 3567$\\
$0.000\leq \Delta (g-r)_\mathrm{P}<0.025$&$1.248\pm0.011$&$0.310\pm0.002$&$-8.217\pm0.044$&$0.062\pm0.001$&$0.071$&$ 3745$\\
$0.025\leq \Delta (g-r)_\mathrm{P}<0.050$&$1.216\pm0.014$&$0.324\pm0.002$&$-8.420\pm0.052$&$0.061\pm0.001$&$0.076$&$ 2602$\\
$0.050\leq \Delta (g-r)_\mathrm{P}<0.075$&$1.208\pm0.019$&$0.345\pm0.003$&$-8.799\pm0.072$&$0.057\pm0.001$&$0.064$&$ 1253$\\
$0.075\leq \Delta (g-r)_\mathrm{P}<0.100$&$1.128\pm0.035$&$0.342\pm0.005$&$-8.575\pm0.135$&$0.067\pm0.003$&$0.072$&$  451$\\
\hline
$\Delta (g-r)_\mathrm{P}<0.02, 0.27\leq C_\mathrm{inv}<0.29$&$0.942\pm0.045$&$0.248\pm0.010$&$-6.381\pm0.215$&$0.099\pm0.004$&$0.103$&$  490$\\
$\Delta (g-r)_\mathrm{P}<0.02, 0.29\leq C_\mathrm{inv}<0.31$&$1.103\pm0.018$&$0.269\pm0.004$&$-7.110\pm0.079$&$0.084\pm0.001$&$0.088$&$ 2356$\\
$\Delta (g-r)_\mathrm{P}<0.02, 0.31\leq C_\mathrm{inv}<0.33$&$1.167\pm0.015$&$0.273\pm0.003$&$-7.311\pm0.056$&$0.077\pm0.001$&$0.083$&$ 2862$\\
$\Delta (g-r)_\mathrm{P}<0.02, 0.33\leq C_\mathrm{inv}<0.35$&$1.260\pm0.019$&$0.281\pm0.003$&$-7.668\pm0.061$&$0.073\pm0.002$&$0.081$&$ 2124$\\
$\Delta (g-r)_\mathrm{P}<0.02, 0.35\leq C_\mathrm{inv}<0.37$&$1.285\pm0.021$&$0.284\pm0.003$&$-7.779\pm0.063$&$0.064\pm0.001$&$0.075$&$ 1599$\\
$\Delta (g-r)_\mathrm{P}<0.02, 0.37\leq C_\mathrm{inv}<0.39$&$1.341\pm0.029$&$0.278\pm0.004$&$-7.774\pm0.090$&$0.064\pm0.002$&$0.076$&$  864$\\
$\Delta (g-r)_\mathrm{P}<0.02, 0.39\leq C_\mathrm{inv}<0.41$&$1.413\pm0.042$&$0.304\pm0.005$&$-8.422\pm0.128$&$0.061\pm0.003$&$0.075$&$  466$\\
$\Delta (g-r)_\mathrm{P}\geq 0.02, 0.27\leq C_\mathrm{inv}<0.29$&$1.214\pm0.048$&$0.323\pm0.009$&$-8.418\pm0.220$&$0.062\pm0.003$&$0.067$&$  257$\\
$\Delta (g-r)_\mathrm{P}\geq 0.02, 0.29\leq C_\mathrm{inv}<0.31$&$1.177\pm0.023$&$0.335\pm0.004$&$-8.543\pm0.096$&$0.062\pm0.001$&$0.068$&$ 1153$\\
$\Delta (g-r)_\mathrm{P}\geq 0.02, 0.31\leq C_\mathrm{inv}<0.33$&$1.208\pm0.022$&$0.335\pm0.003$&$-8.618\pm0.075$&$0.063\pm0.001$&$0.069$&$ 1308$\\
$\Delta (g-r)_\mathrm{P}\geq 0.02, 0.33\leq C_\mathrm{inv}<0.35$&$1.177\pm0.024$&$0.327\pm0.003$&$-8.401\pm0.085$&$0.062\pm0.002$&$0.070$&$ 1028$\\
$\Delta (g-r)_\mathrm{P}\geq 0.02, 0.35\leq C_\mathrm{inv}<0.37$&$1.106\pm0.027$&$0.333\pm0.004$&$-8.358\pm0.096$&$0.059\pm0.003$&$0.068$&$  747$\\
$\Delta (g-r)_\mathrm{P}\geq 0.02, 0.37\leq C_\mathrm{inv}<0.39$&$1.269\pm0.040$&$0.332\pm0.005$&$-8.688\pm0.137$&$0.061\pm0.003$&$0.071$&$  337$\\
$\Delta (g-r)_\mathrm{P}\geq 0.02, 0.39\leq C_\mathrm{inv}<0.41$&$1.305\pm0.064$&$0.336\pm0.007$&$-8.834\pm0.213$&$0.061\pm0.005$&$0.074$&$  163$\\
\enddata
\tablecomments{$a$, $b$, and $c$ are coefficients of the FP (Equation \ref{eq:fp}), while $\varepsilon$ is the intrinsic scatter in the FP (see Equation \ref{eq:chi}). The rms is observed root-mean-square values. $N$ is the number of ETGs in each category.
}
\end{deluxetable*}

\end{document}